\documentclass[
 twocolumn,
superscriptaddress,
 amsmath,amssymb,
 aps,
 pra,
 floatfix,
]{revtex4-2}
\usepackage[normalem]{ulem}
\usepackage{graphicx}
\usepackage{dcolumn}
\usepackage{bm}
\usepackage[colorlinks]{hyperref}
\hypersetup{%
	plainpages=true,
	breaklinks=true,
	hypertexnames=false,
	pageanchor=true,
	colorlinks=true,
	linkcolor={blue},
	citecolor={red},
	urlcolor={blue},
	anchorcolor={black}
}

\usepackage{physics}
\usepackage{xcolor}
\usepackage{soul}
\usepackage{mleftright}
\usepackage{float} 
\newcommand{\sz}{\hat \sigma_z}

\newcommand{\sx}{\hat \sigma_x}

\newcommand{\aop}{\hat a}
\newcommand{\adop}{\hat a ^\dagger}
\newcommand{\opa}{\hat a}
\newcommand{\opad}{\hat a ^\dagger}
\newcommand{\opb}{\hat b}
\newcommand{\opbd}{\hat b ^\dagger}
\newcommand{\opx}[1]{\hat {#1}}
\newcommand{\opxd}[1]{\hat {#1} ^\dagger}
\newcommand{\Tket}[2]{\ket{\tilde{#1},\tilde{#2}}}

\renewcommand{\ket}[1]{|{#1}\rangle}  
\renewcommand{\braket}[2]{\langle{#1}|{#2}\rangle}

\newcommand{\brakket}[3]{\langle{#1}|{#2}|{#3}\rangle}
\newcommand{\expec}[1]{\left\langle{#1}\right\rangle}

\newcommand{\figref}[1]{\mbox{Fig.~\ref{#1}}}
\newcommand{\figuref}[1]{\mbox{Figure~\ref{#1}}}

\newcommand{\figrefs}[1]{\mbox{Fig.s~\ref{#1}}}
\newcommand{\figurefs}[1]{\mbox{Figures~\ref{#1}}}

\newcommand{\secref}[1]{\mbox{Sec.~\ref{#1}}}

\newcommand{\appref}[1]{\mbox{Appendix~\ref{#1}}}
\renewcommand{\eqref}[1]{\mbox{Eq.~(\ref{#1})}}
\newcommand{\equref}[1]{\mbox{Equation~(\ref{#1})}}

\newcommand{\bea}{\begin{eqnarray}}
\newcommand{\eea}{\end{eqnarray}}

\usepackage{xcolor}

\definecolor{darkgreen}{rgb}{0.1, 0.3, 0.0}

\newcommand{\figpanel}[2]{Fig.~\hyperref[#1]{\ref*{#1}(#2)}}
\newcommand{\figpanels}[3]{Fig.~\hyperref[#1]{\ref*{#1}(#2)-(#3)}}
\newcommand{\figpanelNoPrefix}[2]{\hyperref[#1]{\ref*{#1}(#2)}}

\usepackage{changes}
\begin{document}

\title{From Few to Many Emitters Cavity QED: \\ Energy Levels and Emission Spectra From Weak to Deep-Strong Coupling}
\author{Andrea Zappal\'a}
\affiliation{Dipartimento di Scienze Matematiche e Informatiche, Scienze Fisiche e  Scienze della Terra, Universit\`{a} di Messina, I-98166 Messina, Italy}

\author{Alberto Mercurio}
\affiliation{Laboratory of Theoretical Physics of Nanosystems (LTPN), Institute of Physics, Ecole Polytechnique Fédérale de Lausanne (EPFL), CH-1015 Lausanne, Switzerland}
\affiliation{Center for Quantum Science and Engineering, EPFL, CH-1015 Lausanne, Switzerland}

\author{Daniele Lamberto}
\affiliation{Dipartimento di Scienze Matematiche e Informatiche, Scienze Fisiche e  Scienze della Terra, Universit\`{a} di Messina, I-98166 Messina, Italy}

\author{Samuel Napoli}
\affiliation{Dipartimento di Scienze Matematiche e Informatiche, Scienze Fisiche e  Scienze della Terra, Universit\`{a} di Messina, I-98166 Messina, Italy}

\author{Omar Di Stefano}
\affiliation{Dipartimento di Scienze Matematiche e Informatiche, Scienze Fisiche e  Scienze della Terra, Universit\`{a} di Messina, I-98166 Messina, Italy}

\author{Salvatore Savasta}
\affiliation{Dipartimento di Scienze Matematiche e Informatiche, Scienze Fisiche e  Scienze della Terra, Universit\`{a} di Messina, I-98166 Messina, Italy}

\date{\today}


\begin{abstract}

We present a systematic study of the properties of systems composed of $N$ two-level quantum emitters coupled to a single cavity mode, for light-matter interaction strengths ranging from the weak to the ultrastrong and deep-strong coupling regimes. 
Beginning with an analysis of the energy spectrum as a function of the light-matter coupling strength, we examine systems with varying numbers of emitters, from a pair to large collections, approaching the thermodynamic limit ($N \to \infty$). 
Additionally, we explore the emission properties of these systems under incoherent excitation of the emitters, employing a general theoretical framework for open cavity-QED systems, which is valid across all light-matter interaction regimes and preserves gauge invariance within truncated Hilbert spaces.
Furthermore, we study the influence of the emitter–environment interaction on the spectral properties of the system. Specifically, when each emitter interacts independently with its own reservoir, we observe the emergence of an emission peak at the cavity’s resonant frequency for even values of $N$.
Our analysis also clarify the evolution of the system as the number of emitters increases, ultimately converging towards an equivalent system composed of two interacting single-mode bosonic fields.

\end{abstract}
\maketitle
\section{Introduction}

Historically, cavity quantum electrodynamics (QED) has centered around two principal interaction regimes: the weak-coupling and the strong-coupling regimes \cite{Vahala2003optical, walther2006cavity}.
In the weak-coupling regime, dissipation dominates over the intrinsic coherent coupling between light and matter. 
This regime is characterized by the Purcell effect \cite{Purcell1946}, which gave rise to several interesting applications as low-threshold solid-state lasers \cite{Vahala2003optical}
and efficient single-photon emitters \cite{senellart2017high, Salter2010entangled, Somaschi2016near}.
In the strong-coupling regime, system losses are sufficiently low with respect to the coupling rate to enable the observation of vacuum Rabi oscillations \cite{raimond2001manipulating}, where energy coherently oscillates between the two-level system (TLS) and the cavity mode. This regime has pioneered the development of second generation quantum technologies \cite{dicarlo2009demonstration, clarke2008superconducting, haroche2013nobel, Kuhn2002, Volz2012}.

If the light–matter interaction strength reaches a non-negligible fraction of the relevant transition frequencies of the components, the system enters the so-called ultrastrong coupling (USC) regime. In this regime, the interaction can significantly change the system properties and, nowadays, it has been achieved in a great variety of systems and settings \cite{Frisk2019ultrastrong, Forn2019ultrastrong}.
A number of intriguing physical effects have been theoretically predicted in the USC regime. Potential applications include fast and protected quantum information processing \cite{nataf2011protected, stassi2020scalable, chen2021fast} and quantum nonlinear optics processes \cite{garziano2015multiphoton, garziano2016one, kockum2017deterministic, wang2024strong}. 
As the light-matter interaction strength increases further, it becomes possible to enter a regime where the coupling strength surpasses the resonance frequencies of the material and the quantized light modes—this is known as the deep strong-coupling (DSC) regime \cite{DeLiberato2017virtual}. One of the most intriguing phenomena predicted in this regime is the effective decoupling of light and matter \cite{Mueller2020deep,Ashida2021, mercurio2022regimes}.
Within these regimes, standard approximations break down, allowing processes that do not conserve the number of excitations in the system, leading, for example, to a ground state that contains virtual excitations  \cite{stassi2013spontaneous, stassi2016output, di2017feynman, de2017virtual}.

The USC and DSC regimes between a single TLS and an  electromagnetic resonator can be described by the the Quantum Rabi Model (QRM) \cite{rabi1936process} or by its simple generalizations to describe parity symmetry breaking and/or additional cavity modes (see, e.g. \cite{niemczyk2010circuit, wang2023probing}).
At our knowledge, these interaction regimes have been reached within an individual quantum emitter only in circuit QED settings, where the interaction strength is not limited by the smallness of the fine-structure constant.
The USC light-matter coupling can also be achieved in a great variety of systems by coupling to light many dipoles or collective excitations as excitons, magnons, plasmons, intersubband and Landau transitions \cite{cacciola2014,baranov2020,ghirri2023,pisani2019,keller2020}.
In these cases, the relevant matter excitations can be approximated by bosonic fields, hence all these systems can be described by the Hopfield-like models, describing coupled quantum harmonic oscillators \cite{Hopfield1958}.

Here we explore the USC and DSC in systems with more than $N=1$ emitter. In systems with a few quantum emitters, cooperative effects can enhance the coupling strength, but, interestingly,  the system still displays anharmonicity.
It has been shown that systems of few quantum emitters coupled to a plasmonic nanocavity can reach the USC regime (see, e.g., \cite{Chikkaraddy2016,ojambati2019,mercurio2022}).
We study these few emitter systems by considering a generalized Dicke model in the multipolar gauge, including the quadratic self-polarization term, usually omitted. This term ensures gauge invariance \cite{DiStefano2019resolution,garziano2020gauge,settineri2021} and determines the convergence to the Hopfield model in the thermodynamic limit.
We also analyze the evolution of this prototypical light-matter system as a function of the number of emitters by studying the energy levels and the emission properties.
Adopting a gauge-invariant master equation \cite{akbari2023generalized}, we calculate emission spectra under incoherent excitation of the emitters, described by coupling the emitters to reservoirs with a given effective temperature, while considering the resonator coupled to a zero-temperature reservoir.

We also study the impact on the spectra of considering a collective reservoir, individual (local) reservoirs, and their co-existence for the ensemble of emitters. In general, when the emitters are all located within a region smaller than a wavelength of the relevant radiation field, their interaction with a collective reservoir is expected to dominate. Local impurities, defects, interaction of quantum emitters as molecules with vibrational degrees of freedom can give rise to decoherence effect which can be better modeled considering individual reservoirs for the emitters \cite{shammah2018open, chase2008collective}. The interaction with a local reservoir can be boosted, for example, in circuit QED systems, coupling each qubit with  an individual microwave antenna.

In Sec. \ref{sec:Theory} we introduce the model Hamiltonians, both in the Coulomb and multipolar gauge for a system of $N$ two-level emitters coupled to  a single-mode resonator via (electric) dipolar interactions. In this section we will also discuss the thermodynamic limit of this model. In Sec. \ref{sec:Energy}, after introducing a scheme for labeling the energy levels of the emitters-cavity system beyond the rotating-wave approximation, we investigate the influence of the transverse self-polarization term $P^2$ on the energy levels. We also present the energy spectra of the emitters-cavity system as a function of the coupling strength from the weak to the DSC regime, considering systems with different number of emitters: $N =1\,$-$\,6, 10, 50, 100$, and, finally, $N \to \infty$. In Sec. \ref{sec:OQS} we introduce the interaction of the system components with their reservoirs. We present the master equation that will be used to calculate the system dynamics and the emission spectra. This section also shows how to calculate the output photon rate and emission spectra. The calculated emission spectra are presented in Sec. \ref{sec:Spectra}, with an analysis of their main features. Conclusions are finally drawn in Sec. \ref{thend}.

%
%
\section{Hamiltonians} \label{sec:Theory}

In this section, we present the theoretical description of systems composed by $N$ non-interacting identical two-level emitters coupled to a single mode electromagnetic resonator. 
Throughout this work, we employ the dipole approximation. The resulting model corresponds to an extended QRM derived by applying the two-level approximation to the full light-matter Hamiltonian.

This framework is suitable for the description of both natural atoms (after a two-level approximation) placed inside a cavity, as well as superconducting qubits interacting with a $LC$ resonator, though some microscopic differences exist between these implementations. Indeed, in the case of natural atoms, as no particular geometry is specified, we assume the emitters to be sufficiently spaced apart such that electrostatic interactions among them can be safely neglected.

\subsection{Finite Number of Emitters} \label{subsec:finite_emit}
The Coulomb Gauge Hamiltonian of a single emitter interacting with the single mode electromagnetic field, is given by \cite{garziano2020gauge, savasta2021gauge}
\begin{equation} \label{eq:HC_1N}
        \begin{split}
        {\cal \hat H}_{\rm C} = \hbar\omega_c\opad\opa  + \frac{\hbar\omega_a}{2} \big[ 
        & \hat{\sigma}_z \cos(2 \lambda(\opad+\opa)) +\\
        & +\hat{\sigma}_y \sin(2 \lambda(\opad+\opa))\big] \, ,
    \end{split}
\end{equation}
where $\opa$ is the photon annihilation operator, 
$\omega_c$ is the frequency of the single mode electromagnetic field, $\omega_a$ is the transition frequency of the TLs, and $\hat \sigma_\alpha$ (with $\alpha=x,y,z$) are the Pauli operators.
The effective collective coupling is defined as $\lambda = \eta\sqrt{N}$, where $\eta$ is the individual normalized coupling strength, defined as the single-dipole coupling strength over the resonator frequency, $\eta = g/\omega_c$. The first term in the Hamiltonian in \eqref{eq:HC_1N} is the free electromagnetic field contribution, while the second one describes the matter contribution including the interaction with light.

\equref{eq:HC_1N} can be extended to describe a system on $N$ two-level electric dipoles \cite{garziano2020gauge}
\begin{equation} \label{eq:HC_multiN}
        \begin{split}
        {\cal \hat H}_{\rm C} = \hbar\omega_c\opad\opa  + \frac{\hbar}{2}\sum_i^N \omega_a^{(i)} \big[ 
        & \hat{\sigma}_z^{(i)} \cos(2 \lambda(\opad+\opa)) +\\
        & +\hat{\sigma}_y^{(i)} \sin(2 \lambda(\opad+\opa))\big] \, .
    \end{split}
\end{equation}
If the $N$ TLs are identical, \eqref{eq:HC_multiN} can be rewritten using collective angular momentum operators, ${\hat J}_{\alpha}= \frac{1}{2}\sum_i^N \sigma_{\alpha}^{(i)}$. Thus, the Hamiltonian reads
\begin{equation} \label{eq:H_C_Dicke}
    \begin{split}
        {\cal \hat H}_{\rm C} = \hbar\omega_c\opad\opa  + \hbar\omega_a \big[ 
        & \hat{J}_z \cos(2 \lambda(\opad+\opa)) +\\
        & +\hat{J}_y \sin(2 \lambda(\opad+\opa))\big] \, .
    \end{split}
\end{equation}
The corresponding multipolar gauge Hamiltonian can be obtained by applying a suitable unitary transformation \cite{garziano2020gauge}, $\hat{ \mathcal{T}} = \exp\left( -2 i \lambda \hat{J}_x \left( \opa + \opad \right) \right)$, to the Coulomb gauge Hamiltonian
\begin{equation} \label{eq:H_D_DickeNi}
\begin{split}
    &\hat{\cal H}_{D} =
     \hbar\omega_c\opad\opa + \frac{\hbar}{2}\sum_{i}^{N}\omega_a^{(i)}\sz^{(i)} \\
    & - i\hbar\lambda\omega_c(\opa-\opad)\sum_{i}^{N}\sx^{(i)} +
    \hbar\lambda^2\omega_c \biggl( \sum_{i}^{N}\sx^{(i)} \biggr)^2 \, .
\end{split}
\end{equation}
Analogously to the Coulomb gauge, if the TLSs are identical, it can be written as
\begin{equation} \label{eq:H_D_Dicke}
    {\cal \hat H}_{\rm D} = \hbar \omega_c \opad \opa + \hbar \omega_a \hat{J}_z + 2 i \hbar \lambda \omega_c \hat{J}_x \left( \opad - \opa \right) + 4 \hbar \lambda^2 \omega_c\hat{J}_x^2 \, .
\end{equation}
Here we used the same symbols for the photon operators in both the multipolar and Coloumb gauges. However, notice that, the creation and destruction photon operators are not gauge invariant. For example, the photon operator $\hat a$ in the Coulomb gauge transforms in the dipole gauge as: ${{\hat a}_{} = \hat{ \mathcal{T}} {\hat a} \hat{ \mathcal{T}}^{\dagger} = {\hat a} + i\lambda\sx}$.
The first two terms in \eqref{eq:H_D_Dicke} describe the bare 
cavity mode and matter energies respectively, while the latter terms represent the interaction, more precisely the direct matter-cavity interaction and the matter-self interaction term (induced by the light-matter interaction) respectively. The unitary transformation is also a gauge transformation {\cite{garziano2020gauge,DiStefano2019resolution}}, and thus, both representations are equivalent, according to the proper definition of gauge invariance for truncated Hilbert spaces \cite{savasta2021gauge}.

Notice that both \eqref{eq:H_C_Dicke} and \eqref{eq:H_D_Dicke} do not include the electrostatic dipole-dipole interaction term \cite{lamberto2025quantum}. Throughout this work, this term is not taken into account since we focus on a system of non-interacting quantum emitters. This term can become negligible if the quantum emitters are sufficiently separated.
The Hamiltonian in \eqref{eq:H_D_Dicke} is an extended version of the well-known Dicke Hamiltonian \cite{dicke1954coherence}, which includes the transverse self-interaction polarization term $\propto \hat{J}_x^2$. In the dipolar gauge, we refer to this term as quadratic self-polarization term ($P^2$), which plays a crucial role in the accurate description of the non-perturbative ultrastrong coupling regime. For convenience in the following we will refer to this extended version of the Dicke model simply as Dicke model, while calling {\em standard} Dicke model the one without the $\hat{J}_x^2$ term.
%
%
%
\subsection{Thermodynamic Limit} \label{subsec:Therm_lim}
In \secref{subsec:finite_emit}, we introduced the \eqref{eq:H_D_DickeNi} to describe systems with a finite number $N$ of emitters interacting with a single mode cavity electromagnetic field. We now extend this study to consider the thermodynamic limit with $N \to \infty$ and the collective coupling ($\eta \sqrt{N} \equiv \lambda$) constant.
We start applying the Holstein-Primakoff mapping \cite{HolsteinPrimakoff}, used to represent the spin operators in terms of bosonic operators ($\opx{b}$ and $\opxd{b}$):
${\hat J}_z = \opxd{b}\opx{b}$ and ${{\hat J}_+ = \opbd\sqrt{2j - \opbd\opb}}$, (and ${{\hat J}_- = {\hat J}_+^{\dagger}}$), where $j$ is the maximum total spin. 
Starting from the Dicke Hamiltonian in the Coulomb gauge [\eqref{eq:H_C_Dicke}], applying the Holstein-Primakoff mapping, and taking the thermodynamic limit, the resulting Coulomb-gauge Hamiltonian \cite{garziano2020gauge,lamberto2025quantum} is
\begin{equation} \label{eq:Hhop_C}
    \begin{split}
     \hat   {\cal H}_C^{(\infty)} &= \hbar\omega_c\opad\opa + \hbar\omega_a\opbd\opb - i\hbar\omega_a\lambda(\opbd - \opb)(\opad + \opa) +\\
        &+\hbar\omega_a\lambda^2(\opad +\opa)^2 \, .
    \end{split}
\end{equation}
It is constituted by the free electromagnetic field contribution $\hbar\omega_c\opad\opa$,  the free bosonic Hamiltonian, originating from the matter  system ($\hbar\omega_a\opbd\opb$), the boson-boson  interaction term, and the diamagnetic term $\propto (\opad +\opa)^2$.
The thermodynamic limit dipole gauge Hamiltonian can be achieved through an analogous approach starting from the \eqref{eq:H_D_Dicke}, obtaining the following
\begin{equation} \label{eq:Hhop_D}
    \begin{split}
        {\cal H}_D^{(\infty)} = & \hbar\omega_c\opad\opa + \hbar\omega_a\opbd\opb + i\hbar\omega_a\lambda(\opad - \opa)(\opbd + \opb) +\\
        &+\hbar\omega_a\lambda^2(\opbd +\opb)^2 \, .
    \end{split}
\end{equation}
We observe that these two Hamiltonians, obtained in the two different gauges in the thermodynamic limit, are equivalent to the Hopfield Hamiltonian for the case of a single mode electromagnetic resonator interacting with a bosonic polarization wave \cite{hopfield1958theory, garziano2020gauge}.

\section{Energy Levels} \label{sec:Energy}

In this section we investigate the energy levels of systems constituted by a single-mode electromagnetic resonator interacting with $N$ (ranging from 1 to many, to $\infty$) two-level quantum emitters, described in the Coulomb gauge by \eqref{eq:H_C_Dicke}, or equivalently, in the dipole gauge by \eqref{eq:H_D_Dicke}.

First, let’s consider the trivial case where there is no interaction between the light and matter systems ($\lambda =0$). In this scenario, the Hilbert space is divided into two separate subspaces, each representing one of the two non-interacting systems. Consequently, the total eigenstates of the system are simply given by the tensor product of the eigenstates of these two subspaces:
$$\ket{\psi}=\ket{\text {matter states}} \otimes \ket{\text{number of photons}}\,.$$
For instance, in a system composed of a single emitter interacting with a single mode of light, each eigenstate can be represented by a ket with two enters.
The first one specifies the ground ($g$) or excited ($e$) state of the emitter, while the second one indicates the number of photons.
In general, for a system of $N$ emitters, the eigenstates of the non-interacting system can be described with $N+1$ slots: the first $N$ labels denote the excitation states of the emitters,  and the last one indicates the photon number. Each eigenstate corresponds to a specific total excitation number $c$, which is the sum of the number of emitters in the excited state (taking values $0,1,2,...$) plus the photon number $n$.
This can be represented as: $$\ket{\underbrace{e(g),\dots, e(g),}_{N-times} n }\,.$$

\subsection{Tavis-Cummings}
For $\lambda \neq 0$ and neglecting the counter-rotating terms in the interaction Hamiltonian (as well the quadratic self-polarization term), one ends with the Tavis Cummings (TC) model \cite{tavis1968exact}. Here we limit to consider the resonant case: $\omega_a^i = \omega_c$, for all the qubits. In this case, the energy eigenstates can be labeled by three quantum numbers (neglecting for the moment degeneracies): 
$\ket{c,j,k}$,
where \emph{c}, represents the bare energies of the system $H_0$; \emph{j}, the total angular momentum of the spin subsystem; \emph{k}, which indicates the energy level: within a given value of $c$, $k' > k$ implies $E_{k'} >E_k$, being $E_k$ the corresponding eigenenergies.
For each \emph{c} there are a certain number of levels $k_j$, labeled by $k$. We have:
\begin{eqnarray}\label{eq:cjk}
    &j_{\rm min} \leq\,\,\, j \leq \frac{N}{2} \nonumber \,,\\
    &0 \leq\,\,\, c \leq \infty \,,\\
    & k_j =
    \begin{cases}
        c+j+\!\frac{2-N}{2}  &\text{if } c \!< j+\!\frac{N}{2} \\
        2j +1  &\text{if } c \! \geq j+\!\frac{N}{2} 
    \end{cases}\quad\text{ with } 1 \leq k \leq k_j\nonumber \,\nonumber \,,
\end{eqnarray}
where $j_{\rm min}$ is the minimum total angular momentum arising from the composition of $N$ spins: $j_{\rm min}=0$ if $N$ is even, or $j_{\rm min}=1/2$ if $N$ is odd.
The total number of states for each \emph{c} is given by
\begin{equation} \label{eq:deg_levels}
    M_c =
    \begin{cases}
        \sum\limits_{j = j_{\rm min}}^{N/2}  
        k_j d_{k_j} \quad\quad &\text{if } c < N \\
        2^N \quad\quad &\text{ if } c\geq N 
    \end{cases} \, ,
\end{equation}
where $d_{k_j}$ indicates the degeneracy of the $k_j$ group of levels, meaning that each $k \in k_j$ is $d_{k_j}$ degenerate.
This number is given by the formula \cite{tavis1968exact}
\begin{equation}
    d_{k_j} = \frac{N! (2j+1)}{(\frac{N}{2}+j+1)!(\frac{N}{2}-j)!} \, .
\end{equation}
The eigenstates can also be represented indicating the degeneracy of the $k$-th level
$$ \ket{c,j,k,d_k} \, .$$
We observe that such degeneracies do not change with interaction strengths $\lambda \neq 0$. Their number $d_k$ can be omitted when not necessary.
\begin{figure}
    \centering
    \includegraphics[width=1\linewidth]{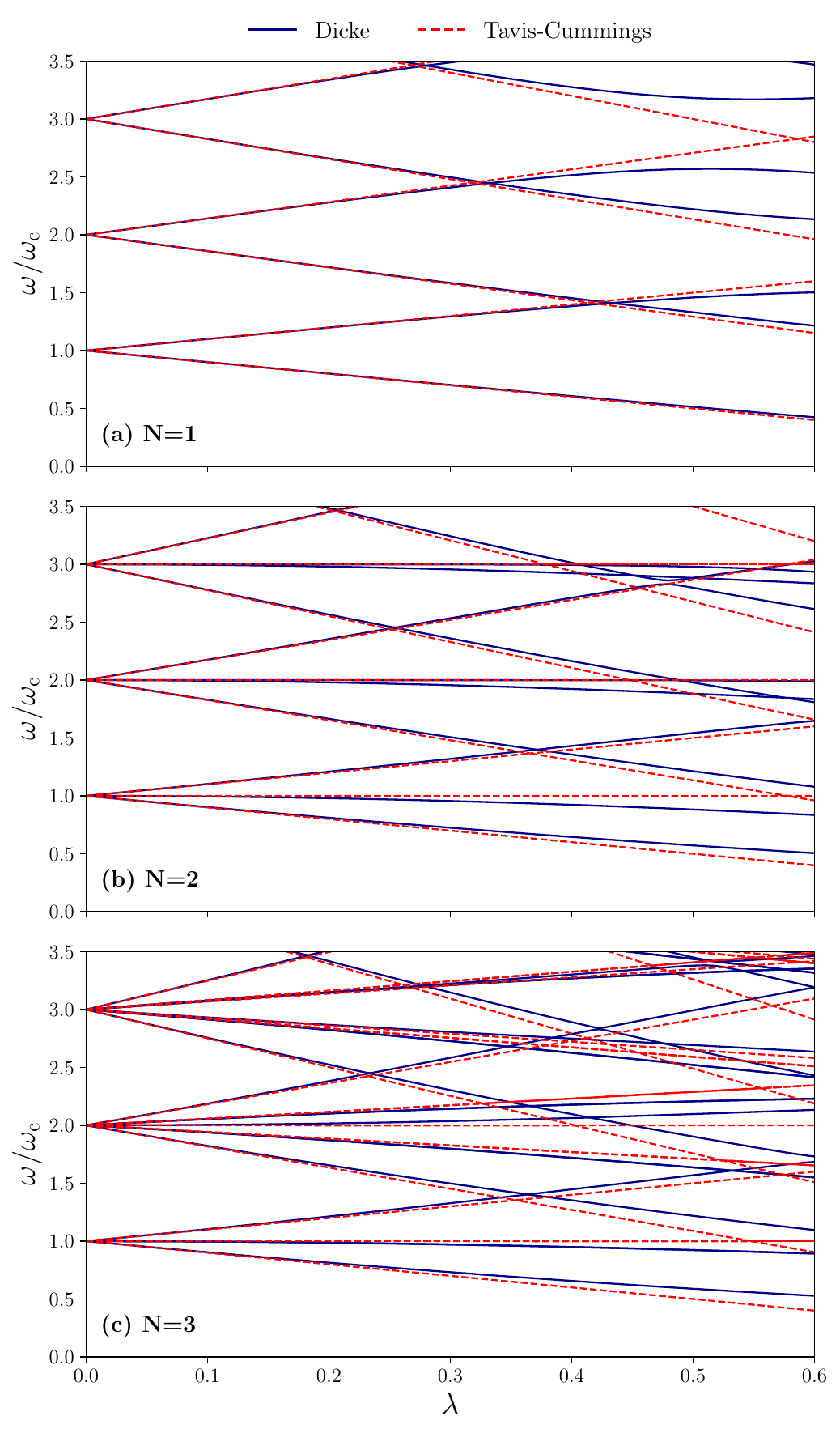}
    \caption{  (a) Comparison between the lowest energy levels  of the QRM (blue continuous line) and those of the JC model, as a function of the normalized coupling strength.
 (b,c) Comparison between the lowest energy levels of the generalized Dicke Hamiltonian in \eqref{eq:H_D_Dicke} (blue continuous line) and those of the TC model (red dashed lines), as a function of the normalized coupling strength for $N=2$ (b) and N=3 (c) TLSs.}
    \label{fig:Tavis_Cummings1}
\end{figure}
While at $\lambda=0$, all the levels with a given $c$ are degenerate (for $\omega_a^i = \omega_c$),
when the interaction between the  two subsystems is switched on, these eigenstates begin to mix, resulting in new eigenstates for the interacting system, partially removing degeneracies.
In the simplest case of a single emitter (JC), all the eigenstates of the non-interacting system, except the ground state, are doubly degenerate at $\lambda=0$. The multiplet with excitation number $c$ is constituted by the eigenstates $\{\ket{c,j =1/2,k}\}$ which splits when $\lambda \neq 0$ as shown in \figref{fig:Tavis_Cummings1}(a), where the eigenenergies are displayed using the ground state ($\ket{0,1/2,1}$) energy as reference.
Higher energy states can be labeled as $\ket{c,1/2,k}$, with $k=1,2$.

\begin{figure}
    \centering
    \includegraphics[width=1\linewidth]{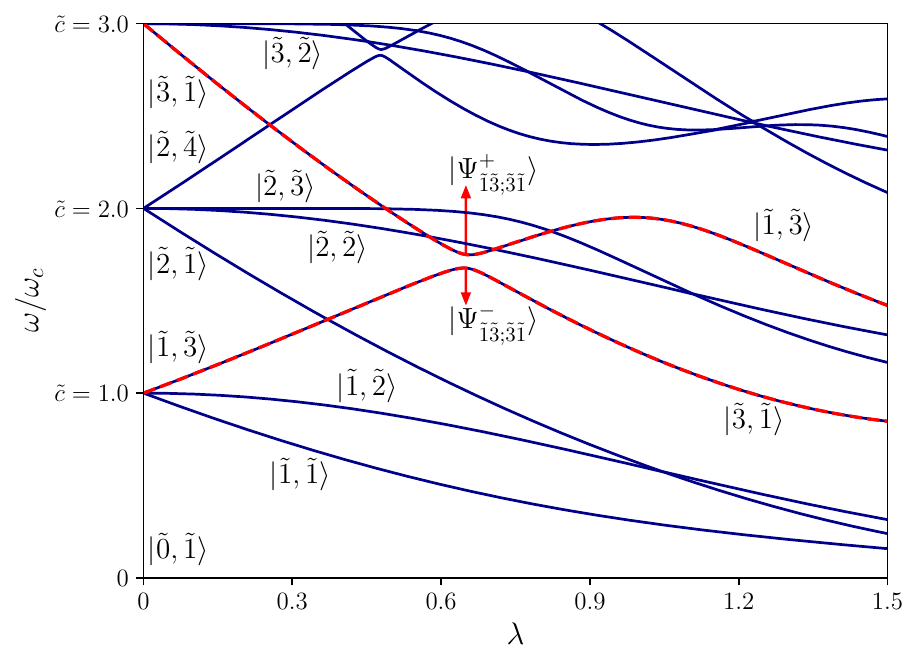}
    \caption{
    Energy states, and their corresponding labeling, of the generalized Dicke Hamiltonian in \eqref{eq:H_D_Dicke} for $N=2$ TLSs, as a function of the normalized coupling strength $\lambda \in \{0,1.5\}$.
    The energy levels are labeled  using the notation  $|\tilde c, \tilde k \rangle$, where $\tilde c$ is the excitation number of the multiplet at $\lambda = 0$ and $\tilde k$ indicates the levels in ascending order of energy within each multiplet.
    Two levels, $\Tket{1}{3}$ and $\Tket{3}{1}$, are highlighted in red to evidence an avoided level crossing occurring at approximately $\lambda \sim 0.65$. This anti-crossing is characterized by the formation of hybridized superposition states, denoted as $\ket{\Psi_{\tilde{c}\tilde{k};\tilde{c'}\tilde{k'}}^{\pm}}$. At the anti-crossing, the energy levels exchange character, and thus, we effectively swap the labels of the two states.}
    \label{fig:levels_labeling}
\end{figure}

For $\lambda  \gtrsim 0.1$, the system enters the USC regime and the counter rotating terms in the Hamiltonian in \eqref{eq:H_D_DickeNi} start to affect both the energy eigenvalues and eigenstates of the system. 
The eigenstates of \eqref{eq:H_D_DickeNi} do not conserve any more the excitation number, and can be written as superpositions of the TC eigenstates with the same parity. These superpositions are induced by the counter-rotating terms in \eqref{eq:H_D_DickeNi}, which can be regarded as a perturbation.
In the following we label the eigenstates of the Dicke model as
\begin{equation}
    \Tket{c}{k} \, .\nonumber
\end{equation}
For each energy level, we identify its label $\tilde{c}$ by looking at its evolution for $\lambda \to 0$, when the Dicke energy levels become indistinguishable from the corresponding TC ones. For example, an energy level which for $\lambda \to 0$ tends to the TC ones described by a given $c$, will have $\tilde c = c$, for any value of $\lambda$.
For each multiplet of given $\tilde c$, we introduce the additional integer number $\tilde{k}$ to label the eigenstates in ascending order of energy, considering the small coupling regime ($\lambda < 0.1$). 
\figuref{fig:levels_labeling} shows an example of labeling for a system of two emitters.
Notice that there is some ambiguity in labeling the energy levels in the vicinity of the anti-crossing regions. When two levels exhibit a small (with respect to the bare resonance frequencies) anticrossing, we label them describing explicitly the superposition at the anticrossing and nearby it. For example, two energy levels $\Tket{c}{k}$ and $\Tket{c'}{k'}$ that mix approaching an anticrossing, are described as $\ket{\Psi^{\pm}_{\tilde{c}\tilde{k};\tilde{c'}\tilde{k'}}}$, where $\ket{\Psi^{\pm}_{\tilde{c}\tilde{k};\tilde{c'}\tilde{k'}}}\simeq \alpha\Tket{c}{k} + \beta_{\pm}\Tket{c'}{k'}$, where $\alpha$ can be chosen to be real and $\beta_\pm = |\beta|\exp(i \theta_\pm)$. Here, $\ket{\Psi^+}$ and $\ket{\Psi^-}$ are the higher and lower splitted energy levels  (red lines in \figref{fig:levels_labeling}). Of course the specific state (including the coefficients $\alpha$ and $\beta$) depends on the specific value of the coupling strength $\lambda$. Outside the anti-crossing region, we choose to label the states as if they would cross. As a result, to the right of the anti-crossing, the labels of the states are exchanged with respect to the left.
Even if not explicitly indicated, the states $\Tket{c}{k}$ always imply a given value of $\lambda$: states calculated using different values of $\lambda$ are different even if they have the same labels. 
Notice that, in contrast to $\tilde c$, the range of values of $\tilde k$ can differ from the range of $k$ values describing the TC levels, owing to the partial removal of degeneracies determined by the counter-rotating terms.  
%
%
\subsection{Self-Energy Polarization Term and Light-Matter Decoupling} \label{subsec:P2_sho}
\begin{figure*}[t]
    \centering
    \includegraphics[width=1\linewidth]{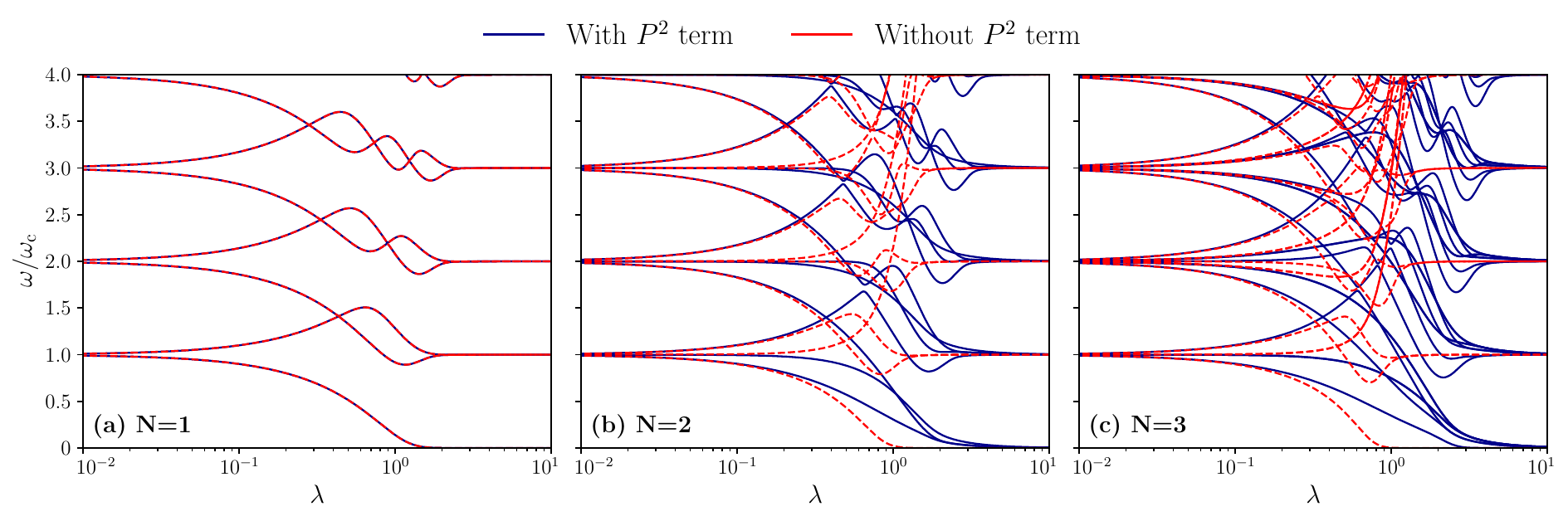}
    \caption{Comparison between the lowest energy eigenvalues of the generalized Dicke Hamiltonian in \eqref{eq:H_D_Dicke} (blue solid lines) and those of the standard Dicke Hamiltonian, without the self-energy polarization term (red dashed lines). The energy levels are shown as a function of the normalized coupling strength on a logarithmic scale for $N=1,2,3$.
    The energy levels are displayed considering the ground state energy as reference.}
    \label{fig:noP2}
\end{figure*}

\figuref{fig:noP2} compares the lowest energy eigenvalues of the Hamiltonian in \eqref{eq:H_C_Dicke} (or equivalently \eqref{eq:H_D_Dicke}) (blue solid lines) and those obtained by dropping the self-energy polarization term $\hbar\lambda^2\omega_c \left( \sum_{i}^{N}\sx^{(i)} \right)^2$ in \eqref{eq:H_D_Dicke} (red dashed lines), corresponding to the energy levels of the standard Dicke model. The energy spectra are taken as a function of the normalized light-matter coupling strength $\lambda$ (defined as $\lambda=g\sqrt N/\omega_c$.) in the range $0 \leq \lambda\leq10$, in a logarithmic scale, for systems composed by $N=1,\, 2,$ and $3$ emitters, displayed using the ground state energy as reference.
\figuref{fig:noP2}  clearly shows the impact of the self-polarization term
on the energy spectrum, in the strong coupling regime and beyond ($\lambda \gtrsim 0.1$).
Notice that, the normalized eigenvalues displayed in \figref{fig:noP2}\,(a) for $N=1$ are not affected by the self-polarization term, since, in the strict two-level approximation employed here, this term is proportional to the identity operator ($\hat{\sigma}_x^2 = \hat{I}$), so it only gives a global constant shift in energy.

\figuref{fig:noP2}(b-c)  show the calculations for $N=2$ (b) and $N=3$ (c) emitters, respectively. In these cases, dropping the self-polarization term, significantly alters the behavior of the energy levels. In the absence of this term, the energy levels diverge and do not approach the matter-decoupled limit as $\lambda \gg 1$, with the exception of a couple of levels for each multiplet, which become degenerate in the DSC regime.
In this regime ($\lambda \gg \omega_a, \, \omega_c$), the energy levels of \eqref{eq:H_D_Dicke} tend to those of a harmonic oscillator: $E_n = \hbar\omega_c n$ (with $n=1,2,3...$). 
In this limit, the system Hamiltonian can be regarded as that of a perturbed displaced harmonic oscillator, where each level is $2^N$ degenerate.
In this DSC regime, the atomic bare energy term ${\cal \hat H}_I' =\hbar \omega_a \hat{J}_z$, can be treated as a small perturbation of the system, when considering a number of atoms which is small with respect to the excitation number $\tilde c$. In this case, the unperturbed Hamiltonian is
\begin{equation}\label{DSC}
{\cal \hat H}_{0}' = 
 \hbar \omega_c \opad \opa + 2 i \hbar \lambda \omega_c \hat{J}_x \left( \opad - \opa \right) + 4 \hbar \lambda^2 \omega_c\hat{J}_x^2\, .
\end{equation}
It can be expressed as ${\cal \hat H}_{0}' = \hbar\omega_c{\hat{\cal A }}^{\dagger}{\hat{\cal A }}$, where the annihilation displaced operator is given by ${\hat{\cal A }} =  {\hat a} + \hat\alpha$, with $\hat \alpha = 2i\lambda\hat{J}_x$. 
Notice that this procedure works fine even if $\hat \alpha$ is an operator because it commutates with both $\hat a$ and $\hat a^\dag$, thus the new operators $\hat{\cal A }$ and $\hat{\cal A }^\dag$ obey the bosonic commutation relations.
We observe that the self-polarization term ($-4\hbar\lambda^2\hat{J}_x^2$) is essential to get the unperturbed Hamiltonian ${\cal \hat H}_{0}'$ corresponding to that of a harmonic oscillator.
Neglecting it, the resulting energy levels do not exhibit such a harmonic-like behavior and most of them would diverge as shown in \figref{fig:noP2} (red lines) \cite{Rokaj2018,Ruggenthaler2023}.

The eigenstates of ${\cal \hat H}_{0}'$ can be obtained by applying the displacement operator $\hat D(-\hat\alpha)= D^\dag(\hat\alpha)$ to
the states $\ket{n,j,m}$ (more details are given in Appendix \ref{app:harmonic}, see  \eqref{eq:appA_ntilde_Dn})
\begin{equation}\label{nuovi}
   \ket{\tilde{n}_m,j,m} = \hat D^\dagger(\hat\alpha) \ket{n,j, m}\, ,
\end{equation}
where $n$ is the (undisplaced) photon number ($\hat a^\dagger \hat a |n \rangle = n  |n \rangle$), $j$ is the total angular momentum quantum number and here $|m \rangle$ is an eigenstate of $\hat J_x$ with eigenvalue $m$. Notice that, differently from usual displacement operators, $\hat \alpha$ is an operator (proportional to the angular momentum operator $\hat J_x$ acting on the matter Hilbert space). Specifically, $\hat\alpha = 2i\lambda\hat J_x$. When acting on the state $\ket{n,j,m}$, it takes the value of the corresponding eigenvalue $2i\lambda m$. So, \eqref{nuovi} can also be written as follows
\begin{equation} \label{eq:ntilde_Dn}
\ket{\tilde{n}_m,j,m} = \hat D^\dagger(2i\lambda m) \ket{n,j, m}\, .
\end{equation}
In the limit $\lambda \gg \omega_a$, we can neglect the perturbation $\hat {\cal H}'_I$, and the states in \eqref{eq:ntilde_Dn} (eigenstates of $\hat {\cal H}'_0$) can be regarded as approximate eigenstates of the total Hamiltonian $\hat H_D$
\begin{equation}
    \hat H_D \ket{\tilde{n}_m,j,m} \simeq \hbar \omega_c \tilde{n} \ket{\tilde{n}_m,j,m}\, . 
\end{equation}

To understand the rather complex evolution of the energy levels of the Hamiltonian in \eqref{eq:H_D_Dicke} as a function of $\lambda$ (see, e.g., \figref{fig:noP2}), it is interesting to compare these levels (and the corresponding eigenstates) in the high and zero $\lambda$ limit.
In the latter case, the energy states are $|n,j,m_z \rangle$ (where $n$ labels the bare Fock states of the resonator, and $m_z$ the eigenvalue of $\hat J_z$) with eigenvalues $\hbar[ \omega_c n  + \omega_a (N/2 + m_z)]$, corresponding to the specific manifold $\tilde c=n+N/2+m_z$. At $\lambda \gg 1$ the manifolds are spanned only by $\tilde n =n$.
This can explain the behavior (see \figref{fig:noP2}) of levels belonging to a given manifold 
$\tilde c=n+N/2+m_z$ (at $\lambda =0$) which, for $\lambda \gg 1$, reach a lower manifold $\tilde n=n$.
As an example, we can consider the case with $N=2$ emitters [see \figref{fig:noP2}(b)]. Let us consider a level that at  $\lambda=0$ is labeled by $n$, $j=1$, $m=1$. It belongs to the manifold $\tilde c = n+2$.
For $\lambda \gg 1$, it will converge  to the lower energy manifold $\tilde n=n$, corresponding to the energy eigenvalue $\hbar \omega_c \tilde n$. 
Analogously, in \figref{fig:noP2}(c), for $N=3$, the levels with a specific $n$ in the manifold $\tilde c$  ($\lambda=0$), with $m_z=\pm 3/2, \pm 1/2$,  will converge   to the  $\tilde n=n$ manifold for $\lambda \gg 1$. In this case, we observe that each $\tilde n$ manifold (for $\lambda \gg 1$) includes levels originating from the upper manifolds up to $\tilde c =n+3$ for $\lambda=0$.
In general, a given manifold $\tilde n$ (at $\lambda\gg 1$) collects levels originating from the manifolds $\tilde n\leq\tilde{c} \leq \tilde n+N$ at $\lambda = 0$.
This behavior can be understood by observing that, for increasing $\lambda$ the influence of the term $\frac{\hbar}{2} \omega_a \hat J_z$ diminishes significantly. Consequently, levels associated with different values of $\{j,m_z\}$ from different manifolds $\tilde c$, tends to degenerate towards the energy levels of a displaced harmonic oscillator.

The evolution of the energy levels for $\lambda \gg 1$ towards the harmonic levels $\hbar \omega_c \tilde n$ could also be explaine by perturbation theory for degenerates subspaces, by evaluating the matrix elements of the perturbation term ${\cal \hat H}_{I}' = \hbar\omega_a{\hat J}_z$ on the eigenstates $\ket{\tilde n_m,j, m}$ of the unperturbed Hamiltonian ${\cal H}_0'$, obatining
\begin{eqnarray} \label{eq:mat_elem}
    & \bra{\tilde n_m,j,m} \hat{J}_z \ket{\tilde n_{m'},j',m'} =\nonumber \\
    & = \bra{n,j,m}D^\dagger(\hat \alpha) \hat{J}_z D(\hat \alpha)\ket{n,j,m'} \delta_{j,j'}\, ,
\end{eqnarray}
which can be diagonalized in each subspace of given $n$, providing the lowest order correction to the displaced harmonic oscillator Hamiltonian $\hat {\cal H}'_0$.

%


\subsection{Few Emitters} \label{sec:few_energy}
\begin{figure*}[t]
    \centering
    \includegraphics[width=1\linewidth]{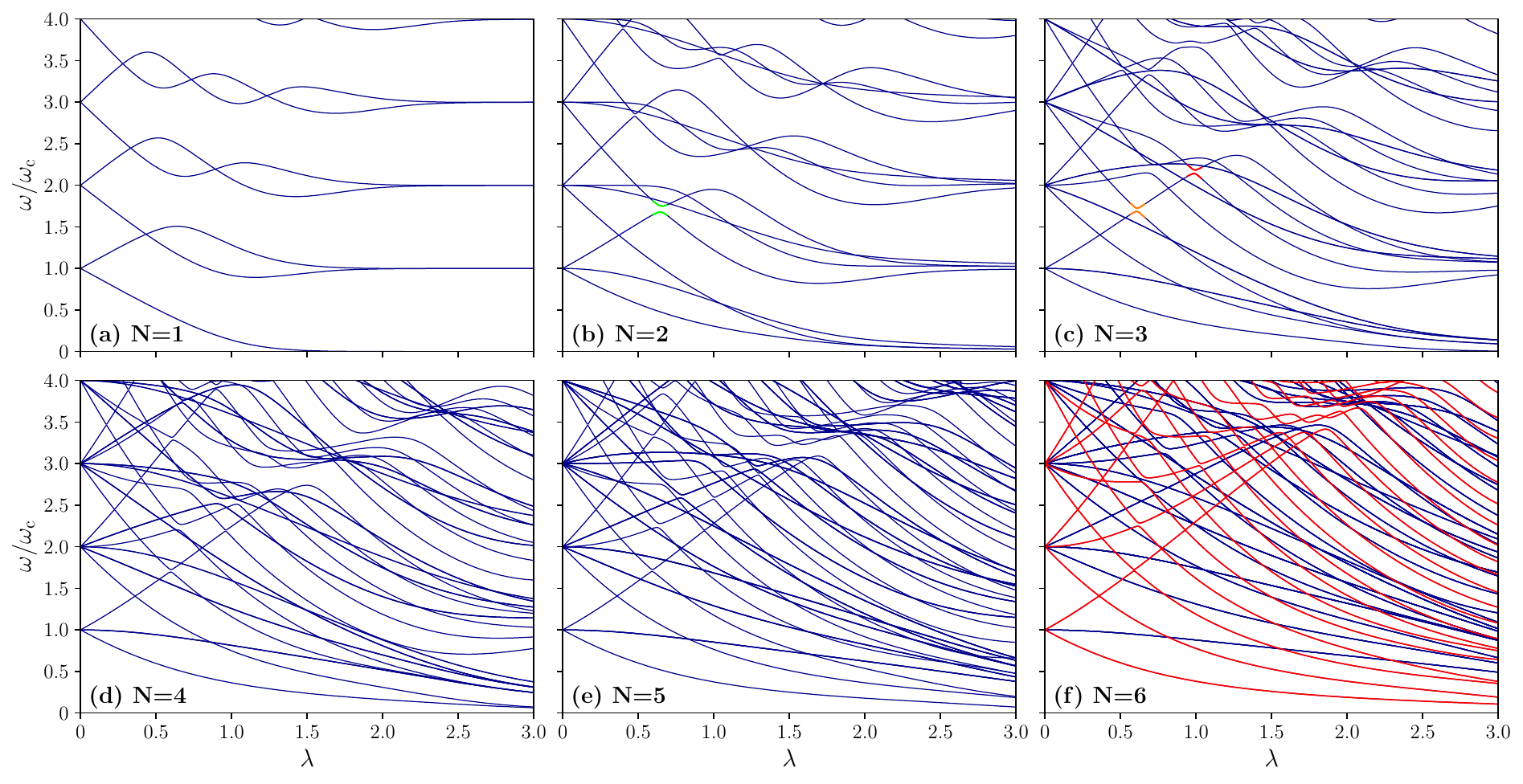}
    \caption{ (a-f) Lowest energy eigenvalues of the generalized Dicke Hamiltonian in \eqref{eq:H_D_Dicke} for systems composed by $N=1$ up to $N=6$ TLSs, varying the normalized coupling strength $\lambda \in \{0,3\}$.
    All the energy levels are taken with the ground state energy as reference.
    Increasing the number of quantum emitters, the number of energy levels rises exponentially as $2^N$.
    Multi-dipole systems present avoided level crossings between energy levels corresponding to states with the same parity, highlighted in green for N=2 (b), and in orange and red for $N=3$ (c).
    These avoided crossings become narrower when the number of TLSs increases.
    (f) The red-highlighted energy levels correspond to the Dicke states belonging to the maximum total angular momentum manifold, $j=N/2$.
    }
    \label{fig:Eigenspectrum_few_all}
\end{figure*}

We now consider the normalized eigenvalues for the Hamiltonian for increasing value of $N$.
Specifically, in \figuref{fig:Eigenspectrum_few_all} are presented  the normalized eigenvalues for the Hamiltonian in \eqref{eq:H_D_DickeNi} with $N$ ranging from $1$ to $6$.
At resonance ($\omega_a = \omega_c$) and in the absence of interaction ($\lambda=0$), we have a harmonic spectrum with values given by $\omega= c \, \omega_c $, with $c\in \mathbb{N}$.  The degeneracy for each level depends on the number of total excitations (number of photons plus the number of excited TLS) and by the number of emitters $N$ and it is equal to $M_{\rm c}$ [given by \eqref{eq:deg_levels}]. 
Observing \figuref{fig:Eigenspectrum_few_all} at increasing  $\lambda$, the energy levels split apart  (WC and SC regimes).

We now examine the single-emitter spectrum in more detail, see \figuref{fig:Eigenspectrum_few_all}(a). The ground state is denoted with $\Tket{0}{1}$. The higher energy multiplets each consist of two degenerate (at $\lambda = 0$) energy levels : $\Tket{c}{1}$ and $\Tket{c}{2}$.
The energy spectrum for $N=2$ emitters is presented in \figref{fig:Eigenspectrum_few_all}(b). The ground state is again  labeled  as $\ket{\tilde{0}, \tilde{1}}$. In this case  the first multiplet, with excitation number $\tilde{c}=1$, comprises three degenerate (at $\lambda =0$)  levels: $\ket{\tilde{1}, \tilde{k}=1,2,3}$. The subsequent multiplets,  for $\tilde{c}\geq2$,  consist of four degenerate (at $\lambda =0$) levels: $\ket{\tilde{c}, \tilde{k}=1,2,3,4}$ [see \eqref{eq:cjk}].
In the case of $N=3$ emitters, shown \figuref{fig:Eigenspectrum_few_all}(c),  the ground state remains $\ket{\tilde{0}, \tilde{1}}$. The first excited multiplet 
($\tilde c=1$) includes three levels $\ket{\tilde{1}, \tilde{k}=1,2,3}$, while the second multiplet ($\tilde c=2$) contains five levels:
 $\ket{\tilde{2}, \tilde{k}=1,\dots,5}$. All higher-energy multiplets, with $\tilde c\geq N=3$, are composed of six levels: $\ket{\tilde{c}\geq3, \tilde{k}=1,\dots,6}$.
An analogous classification procedure can be extended to for $N=4,5,6$ systems [see \figrefs{fig:Eigenspectrum_few_all}~(d-f)].
We now focus on the energy levels behavior as the normalized coupling strength $\lambda$ increases.  In \figref{fig:Eigenspectrum_few_all}(a), we observe that the spectrum consists of pairs of levels $\{\Tket{c}{1},\Tket{c}{2}\}$ which, for $\lambda <1$,
exibit the characteristic ladder structure of the Jaynes-Cummings (JC) modelor Tavis-Cummings for multi-emitters systems.   
However, as $\lambda$ increases [see \figref{fig:Eigenspectrum_few_all}],  or as the number of emitters grows [see \figrefs{fig:Eigenspectrum_few_all}(b-f)] the energy level structure deviates significantly from the predicted of the JC (or TC) model as previously shown in \figref{fig:Tavis_Cummings1}. 
In the single emitter case ($N=1$) [\figref{fig:Eigenspectrum_few_all}(a)], the split energy levels begine to intertwine with adjacent levels originating from the nearest neighboring multiplets.
When a second emitter is added ($N=2$), as shown in \figrefs{fig:Eigenspectrum_few_all} (b), leads to more frequent level crossings and,  in particular, it reveals clearer signatures of  anti-crossing behavior of the system. Those 
are indicative of higher-order interactions which occur specifically between levels of the same parity.

A detailed analysis of \figref{fig:Eigenspectrum_few_all}~(b) corresponding to the case of $N = 2$ emitters, reveals  an avoided level crossing around $\lambda =0.65$ (green lines).  
Specifically, this involves the highest level of the $\tilde c=1$ multiplet, $\Tket{1}{3}$, which interacts with the lowest level  of the $\tilde{c} = 3$ multiplet, $\Tket{3}{1}$.
Similarly, in the $N = 3$ case shown \figuref{fig:Eigenspectrum_few_all}\,(c), two avoided level crossings involving  $\Tket{1}{3}$ are observed: the first, marked in orange, mirrors the behavior seen for $N=2$ emitters, involving $\Tket{1}{3}$ and $\Tket{3}{1}$ ; the second, characterized by a narrower energy gap, marked in red, involves   $\Tket{3}{3}$. 
This latter crossing results from a preceding, broader avoided crossing between $\Tket{3}{3}$ and the lowest level of the $\tilde{c} = 5$ multiplet, $\Tket{5}{1}$.
This pattern persists as the number of emitters increases.  Notice that the avoided crossings become progressively narrower. 
In \figref{fig:Eigenspectrum_few_all}\,(c) additional smaller avoided level crossings are present, the most notable of which involves  $\Tket{2}{3}$ and $\Tket{4}{1}$.
These avoided level crossings are higher-order processes arising from interaction  terms that do not conserve the number of excitations, enabling coupling between lstates belonging to different excitation subspaces.
\figuref{fig:Eigenspectrum_few_all} shows systems up to $N=6$ emitters. It is evident that the features observed for smaller systems persist as the number of emitters increases. 
In particular, for the $N=5$ and $N=6$ cases in \figref{fig:Eigenspectrum_few_all}\,(e-f), the energy gaps at the avoided level crossings become extremely small,so much so that the levels appear to cross, effectively behaving as if the crossings were exact (see also \secref{subsec:Many_energy}.

As discussed in \secref{subsec:P2_sho}, in the DSC regime, when $\lambda \geq 1$, the spectrum of a light-matter system composed of a finite number of quantum emitters undergoes a decoupling.  In such a regime (DSC), each eigenvalue corresponds to that of a displaced harmonic oscillator, with degeneracy $2^N$ [see \eqref{DSC}]. Hence, for this regime, we have 
equally spaced energy levels.

These levels exhibit multiple degeneracies, originating from different manifolds classified by energy expression $\hbar[\omega_c n + \omega_a (N/2+m_z)]$ for $\lambda \ll 1$, as illustrated in \figref{fig:noP2}. In the figure, the energies of the excited states are calculated relative to the ground state energy, and the system is considered in the resonant case, where there is no detuning ($\omega_c = \omega_a$).

More specifically, each eigenstate within the manifold labeled by $\tilde{n}$ (which, for simplicity, we denote as $n$, since $\tilde n = n$) at $\lambda \gg 1$, corresponds to energy levels originating from the range of multiplets $\tilde{c}$ satisfying ${n \leq \tilde{c} \leq n + N}$.
This behavior is clearly illustrated in \figref{fig:Eigenspectrum_few_all}, which shows the cases of $N=1,2,\dots,6$ quantum emitters. 
As $N$ increases, the number of contributing multiplets grows accordingly, reflecting the richer structure of each manifold in the deep strong coupling regime.

In particular, in the case of a single quantum emitter $N=1$, the ${n} = 0$ manifold, the ground state of the displaced harmonic oscillator, includes levels from $\tilde{c}|_{\lambda = 0} = 0, 1$, namely $\Tket{0}{1}$ and $\Tket{1}{1}$ (following the notation $\Tket{c}{k}$).
The ${n} = 1$ manifold includes levels from $\tilde{c}|_{\lambda = 0} = 1, 2$, labeled as $\Tket{1}{2}$ and $\Tket{2}{1}$.
This pattern continues for higher manifolds, consistently reflecting the structure imposed by the coupling in the deep strong coupling regime.

As specific case we look at \figref{fig:Eigenspectrum_few_all}(b), corresponding to the case of $N=2$ emitters, each manifold labeled by ${n}$ consists of four degenerate eigenvalue levels. These levels originate from the multiplets with excitation numbers in the range  $n\leq\tilde{c}\leq n+2$. 
For example,the ${n}=0$ multiplet includes levels from the $\tilde{c}|_{\lambda = 0} = 0,1,2$ multiplets, namely: $\Tket{0}{1},\Tket{1}{1},\Tket{1}{2}$ and $\Tket{2}{1}$.
The next manifold,  ${n}=1$, contains levels from the  $\tilde{c} = 1,2,3$ multiplets: $\Tket{1}{3},\Tket{2}{2},\Tket{2}{3}$ and $\Tket{3}{1}$].

This behavior continues increasing the number $N$ of emitters, as shown in \figref{fig:Eigenspectrum_few_all}(c–f). Notice that the structure of each manifold becomes increasingly complex due to the exponential growth of the Hilbert space dimensions. Specifically, each manifold ${n}$ contains $2^N$ degenerate eigenvalue levels, and these levels originate from a broader range of excitation subspaces, with $\tilde{c}$ spanning from $n$ to $n + N$ at $\lambda = 0$.

In summary, increasing the number of emitters enriches the internal structure of each manifold, reflecting the growing complexity of the system's energy spectrum.

We conclude this subsection observing that
the red energy levels in \figref{fig:Eigenspectrum_few_all}(f) correspond to the eigenvalues with the highest total angular momentum quantum number, $j = N/2$.
It is well established  (see, e.g., Ref.~\cite{Emary2003}) that, when dealing with a larger number of emitters, 
the collective behavior can be described using angular momentum operators. 
Importantly, in the limit of large $N$, the subspace with the highest total angular momentum dominates the light-matter interaction, as it exhibits the strongest effective coupling to the photon field. This allows for a significant reduction in the dimensionality of the Hilbert space, enabling more efficient modeling of the system.

In the following, we will focus on the energy levels associated exclusively with the $j = N/2$ subspace, neglecting states with $j < N/2$ on the emission spectra.



\subsection{Many Emitters and Thermodynamic Limit} \label{subsec:Many_energy}

\begin{figure}[t]
    \centering
    \includegraphics[width=0.96\linewidth]{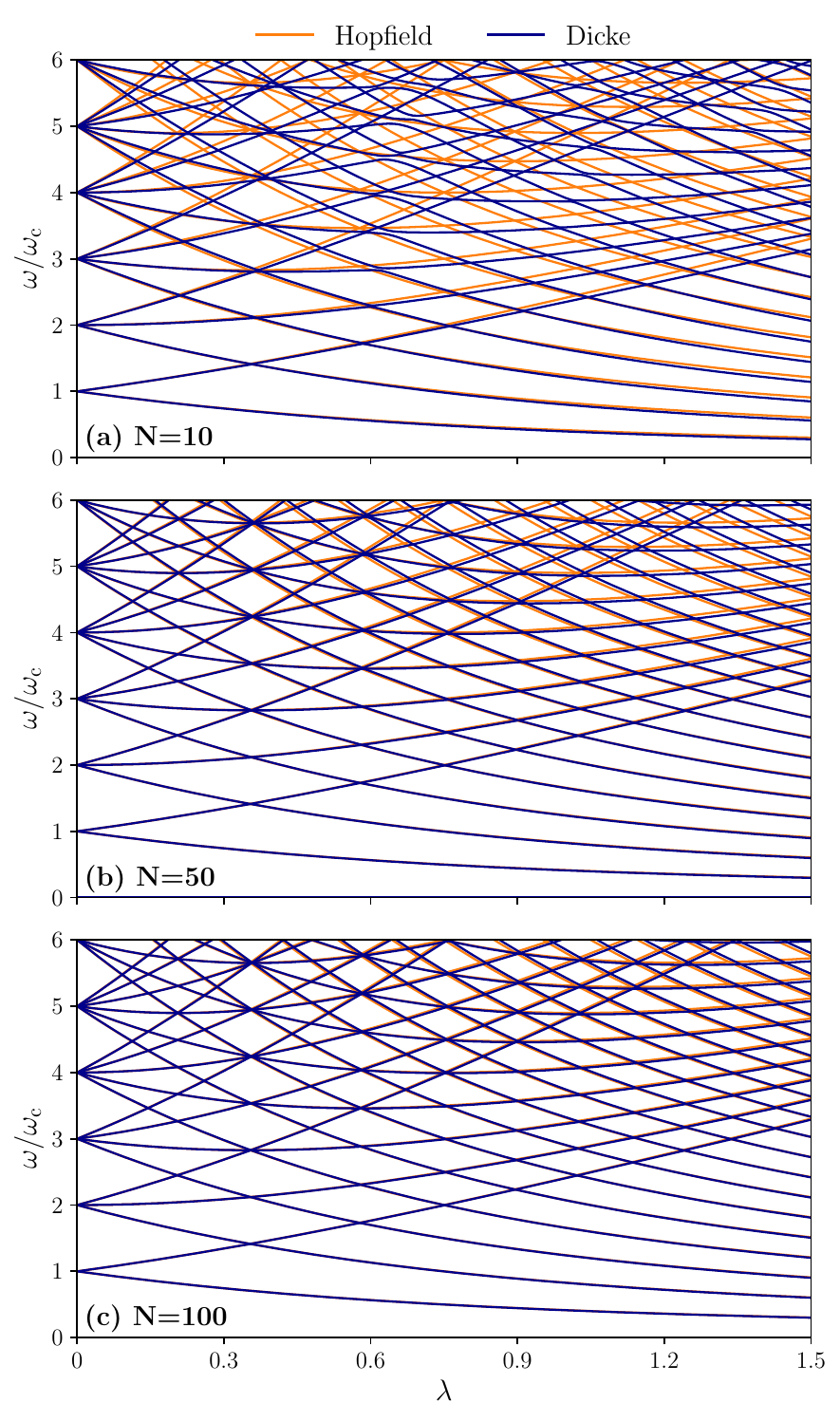}
    \caption{Lowest energy eigenvalues of the generalized Dicke Hamiltonian belonging to the maximum angular momentum manifold (blue solid lines) for $N=10$ (a), $N=50$ (b), and $N=100$ (c), as a function of the normalized coupling strength. For comparison, the corresponding energy levels of the Hopfield model are also displayed (orange solid lines).
    }
    \label{fig:Eigenspectrum_MANY}
\end{figure}

After discussing the eigenvalues for systems with a few emitters, it is interesting to explore the behavior of the energy levels when the number $N$ of emitters grows, until it reaches the thermodynamic limit ($N \rightarrow\infty$, with $\lambda = \eta \sqrt{N}$ constant).
\figuref{fig:Eigenspectrum_MANY} displays the energy levels as a function of $\lambda$ for $N = 10, 50, 100$ emitters (blue curves), considering the ground state energy as reference. 
These eigenvalues are associated with the manifold of maximal total spin, defined by $j=N/2$.
Indeed, in the thermodynamic limit, only the levels within this manifold significantly contribute to the system's dynamics \cite{Emary2003} (see \secref{subsec:Many_spectra}). 
For reference, \figref{fig:Eigenspectrum_MANY} also shows the energy eigenvalues obtained in the thermodynamic limit (orange curves), corresponding to the eigenvalues of the Hopfield model in \eqref{eq:Hhop_D} \cite{Hopfield1958}.
We observe that, already for $N=10$, the energy eigenvalues are quite close to those obtained in the thermodynamic limit, at least for the lowest energy levels. As expected the discrepancies increase at increasing values of the coupling $\lambda$ and at increasing values of $\tilde c$. In particular, the energy levels for $N=10$ still display a number of small avoided crossings, in contrast to the orange curves. For $N=50$, [see \figref{fig:Eigenspectrum_MANY}(b)], the avoided level crossings are no more visible on the adopted scale. The energy levels get very close to the ones in the thermodynamic limit, particularly the ones at lower energy. For small values of $\lambda$ (corresponding to the WC and SC regimes) all the dispalyed energy levels almost do coincide with the corresponding levels in the thermodynamic limit. Some discrepancies at higher energy are still present but less evident respect to the previous case. For $N=100$ [\figuref{fig:Eigenspectrum_MANY}(c)] the eigenvalues of the two models almost coincide for the displayed energy range. However, at higher values of $\tilde c$ (not shown), and hence of energy, the two models will start to differ. 

\section{Open Quantum Systems} \label{sec:OQS}
\subsection{Master Equation} \label{subsec:Master_eq}

We now study the interaction of the system presented in the previous section with external thermal baths.
We describe the  dynamics of the resulting open quantum system by employing the generalized master equation (GME) formalism
\cite{Settineri2018dissipation} (see \appref{sec:GME} for details).
In this framework, the reduced density operator of the system, $\hat{\rho} (t)$,  which is obtained by tracing out the baths degrees of freedom, satisfies the equation
\begin{equation} \label{eq:GME}
    \frac{d}{dt} \hat{\rho}(t) = {\cal L}_{\rm gme}(t) \hat{\rho}(t) \, ,
\end{equation}
where ${\cal L}_{\rm gme}(t)$ is the corresponding GME Liouvillian. The operators entering in $ {\cal L}_{\rm gme}\hat{\rho}$ are
\begin{equation}
\begin{split}
    \hat{S}_i(t) 
    & = \sum_{\epsilon' - \epsilon = \omega} \hat{\Pi}(\epsilon) (\hat{s}_i + \hat{s}_i^{\dagger})  \hat{\Pi}(\epsilon') e^{-i \omega t} \\
    & \equiv \sum_{\epsilon' - \epsilon = \omega} \hat{S}_i(\omega)e^{-i \omega t} \, ,
    \end{split}
\end{equation}
where $\hat{\Pi}(\epsilon)$ is the projector onto the system eigenspace $\ket{\epsilon}$ with associated energy $\epsilon$, and $\hat{s}_i$ are system operators describing the system-bath interaction. 
The operators $\hat{S}_i(t)$ can be decomposed  into positive and negative frequency components, $\hat{S}_i(\omega) \equiv \hat{S}_i^{(+)}(\omega)$ for $\omega>0$ and $\hat{S}_i(\omega) \equiv \hat{S}_i^{(-)}(\omega)$ for $\omega <0$ respectively, associated to the transitions to higher or lower energy eigenstates. 
These operators determine the emission properties of the system. 
In the case of two-level systems, the operators in the Liouvillian are  $\hat s_{\rm ind,\,k}= \hat \sigma_k^-$ for the dissipation of the individual $k$-th quantum emitter, and $\hat s_{\rm coll} = \frac{1}{\sqrt{N}}\sum_k \hat \sigma_k^-$ when considering a collective dissipation channel. 

\subsection{Photon Emission} \label{subsec:Emission}
The  photon emission rate of the system is usually obtained considering the expectation value ${\expec{\adop\aop}_t\equiv\tr{\opad\opa\hat{\rho}(t)}}$. However, this formula is only valid in the WC and SC regimes, when the counter-rotating terms in the interaction Hamiltonian can be neglected.
As the coupling strength increases and the system enters the USC and DSC regimes, the RWA can not be used, and it has been shown that the above formula no longer describes the correct photon emission rate \cite{Ridolfo2013}, due to the influence of the counter-rotating terms. As a consequence, a proper generalization of the input-output theory is needed \cite{di2018photodetection}.

The electric field operator in the Coulomb gauge is
\begin{equation}
    {\hat E} = i\omega_c A_0 (\opa - \opad) \, ,
\end{equation}
where $\opa$ and $\opad$ are the destruction and creation operators in the Coulomb gauge. Therefore, the electric field operator can be decomposed in positive (${\hat E}^+$) and negative (${\hat E}^-$) frequencies as
\begin{equation}
    {\hat {E} ^ {+}} = i\sum_{k>j} \brakket{j_{\rm C}}{(\opa - \opad)}{k_{\rm C}}\ket{j_{\rm C}}\bra{k_{\rm C}} \, ,
\end{equation}
where $\ket{k_{\rm C}}$ is the $k$-th eigenstate in the Coulomb gauge, and $ {\hat {E} ^ {-}}=  ({\hat {E} ^ {+}})^\dag$.
The output photon emission rate is now expressed as a function of the positive and negative frequencies of thee electric field operator, and is given by ${W(t) = \expec{{\hat {E} ^-}{\hat {E} ^+}} \equiv \Tr{{\hat {E} ^-}{\hat {E} ^+}{\hat \rho (t)}}}$ \cite{DiStefano2018,le2020theoretical}.
In the Coulomb gauge, the electric field is proportional to the canonical momentum operator ${\hat \Pi}_{\rm C} = - \varepsilon_0 \hat E$, where $\varepsilon_0$ is the vacuum permittivity.
Consequently, in this gauge,  it is straightforward to express $W(t)$ in terms of canonical variables.
Moreover, $W(t)$ is gauge-invariant, since it represents a physical quantity. Indeed, in the dipole gauge, the electric field operator can be written as ${{\hat E} = - {\hat \Pi}_{\rm D/}\varepsilon_0 - {\hat P}/\varepsilon_0}$, where ${\hat \Pi}_{\rm D}$ is the associated canonical momentum operator.
The unitary gauge transformation operator linking the two previous gauges is
 ${{\hat T} = \exp(-2 i\lambda {\hat J}_x(\aop + \adop))}$ \cite{savasta2021gauge},
obtaining
\begin{equation}
    {\hat E} = i\omega_c A_0 (\opa_{\rm D} - \opad_{\rm D}) \, ,
\end{equation}
where $ \opa_{\rm D} =  {\hat T} {\opa} {\hat T}^{\dagger} = \opa + i\lambda \hat J_x$.
Similarly, the eigenstates transform as
$\ket{j_{\rm D}}={\hat T} \ket{j_{\rm C}} $.
Thus, the positive-frequency component of the electric field operator is
\begin{equation} \label{eq:ElectricField}
    {\hat {E} ^ {+}} = i\sum_{k>j} \brakket{j_{\rm D}}{(\opa_{\rm D} - \opad_{\rm D})}{k_{\rm D}}\ket{j_{\rm D}}\bra{k_{\rm D}} \, .
\end{equation}
\equref{eq:ElectricField} can be re-written by using the Thomas-Reiche-Kuhun (TRK) sum rule  \cite{savasta2020thomas}, as
\begin{equation}
    {\hat {E} ^ {+}} = i\sum_{k>j} \frac{\omega_{kj}}{\omega_c} \brakket{j_{\rm D}}{(\opa_{\rm D} + \opad_{\rm D})}{k_{\rm D}}\ket{j_{\rm D}}\bra{k_{\rm D}} \, ,
\end{equation}
where $\omega_c$ is the cavity frequency and {$\omega_{kj} = \omega_k - \omega_j$} is the transition frequency.
The emission spectrum of the system is evaluated from the steady-state density matrix $\hat \rho_{\rm ss}$ by applying the quantum regression theorem \cite{gardiner2004quantum, napoli2024circuit, mercurio2022regimes}. It is proportional to the two-times correlation function for the electric-field operator
\begin{equation}
    S(\omega) \propto {\rm Re} \int_0^{\infty} d\tau\,e^{-i\omega\tau} \langle \hat E^-(t+\tau) \hat E^+(t) \rangle_{\rm ss} \, .
\end{equation}

\section{Emission Spectra} \label{sec:Spectra}

We begin by analyzing  the system  under incoherent excitation of the quantum emitters. We describe their continuous-wave incoherent pumping by assuming that their thermal bath (which we consider as ohmic) is at an effective temperature $\tilde T_a > \tilde T_c$, , where the $\tilde{T}$ is the normalized temperature by the cavity frequency $\tilde{T} = T/\omega_c$ (with $\hbar = k_B = 1$). In particular, we assume for all the numerical calculations performed in this work $\tilde T_a = 0.15$,
while the cavity temperature is $\tilde T_c = 0$. The emission spectra are calculated as a function of the normalized coupling strength $\lambda$, ranging from the WC to the DSC regimes. 

Each system is initially studied by considering a single dissipation channel for the cavity and a collective dissipation channel for the emitters.
Subsequently, the same system is placed in a configuration where each emitter is coupled to an individual (local) dissipation channel.
Then, we consider a more general configuration, where the emitters are under the influence of both collective and individual reservoirs  (for further details, see Appendix \ref{sec:GME}).
 
For the numerical calculations, we use the following bare damping rates: $\gamma_{\rm c} = 10^{-3} \omega_c$, $\gamma_{\rm coll} = 10^{-3} \omega_c$ (for the collective qubits reservoir), $\gamma_{\rm ind} = 10^{-3} \omega_c$ (for the individual qubits reservoirs). In cases where both types of reservoirs (individual and collective) were present, the dissipation rates $\gamma_{\rm ind}$ and $\gamma_{\rm coll}$ have been halved for a better comparison.


\begin{figure}[b]
    \centering
    \includegraphics[width=1\linewidth]{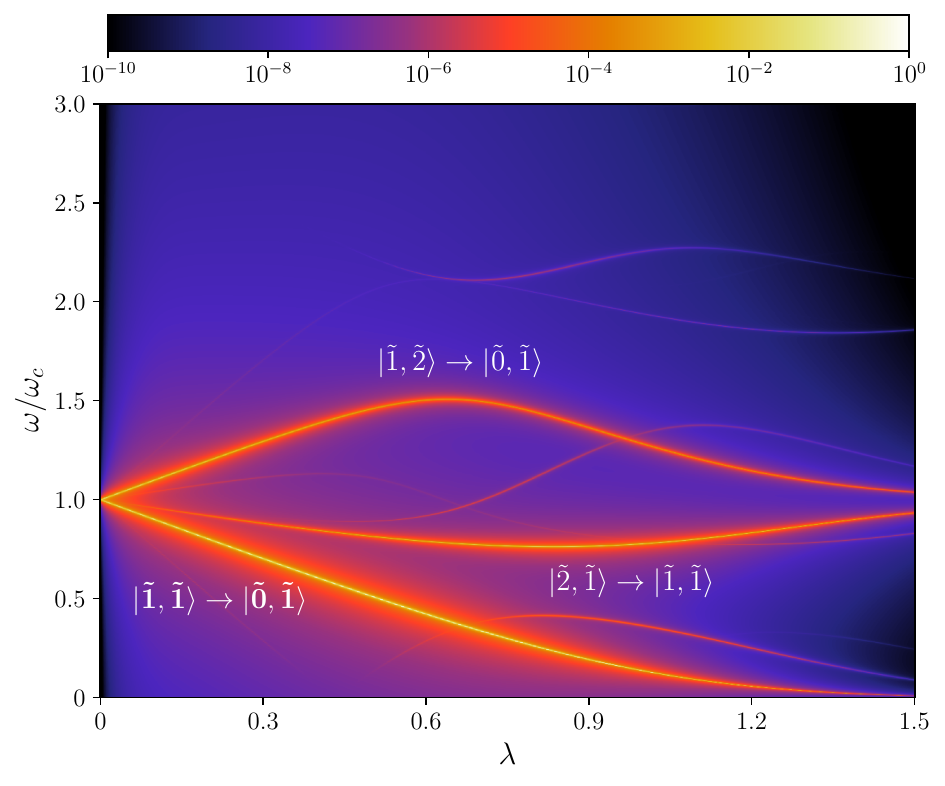}
    \caption{Emission Spectra for a system with a single emitter as a function of the normalized coupling strength. The quantum emitter is considered at an effective temperature of $\tilde T_a = 0.15$.} 
    \label{fig:1atom_spectra}
\end{figure}

\subsection{Single Emitter} \label{subsec:one_spectra}

\figuref{fig:1atom_spectra} shows the simplest case of a single emitter interacting with a single-mode electromagnetic field within the cavity, a well-studied system in literature \cite{mercurio2022regimes, Salmon2022gauge, savasta2021gauge, Settineri2021gauge}. It is included here for a easier comparison with the multi-emitter systems. In this case, the individual and collective dissipation channels coincide.

The eigenstates are labeled as $\Tket{c}{k}$, following the nomenclature introduced in \secref{sec:Energy}, by identifying the manifold $\tilde{c}$ and level $\tilde{k}$, from the lowest to the higher.
The brightest emission lines  in \figref{fig:1atom_spectra} correspond to the transitions $\Tket{2}{1}\rightarrow\Tket{1}{1}$, $\Tket{1}{1}\rightarrow\Tket{0}{1}$ and $\Tket{1}{2}\rightarrow\Tket{0}{1}$.
Additional weaker emission lines involving higher energy levels can also be observed.

%
\begin{figure}[b]
    \centering
    \includegraphics[width=1\linewidth]{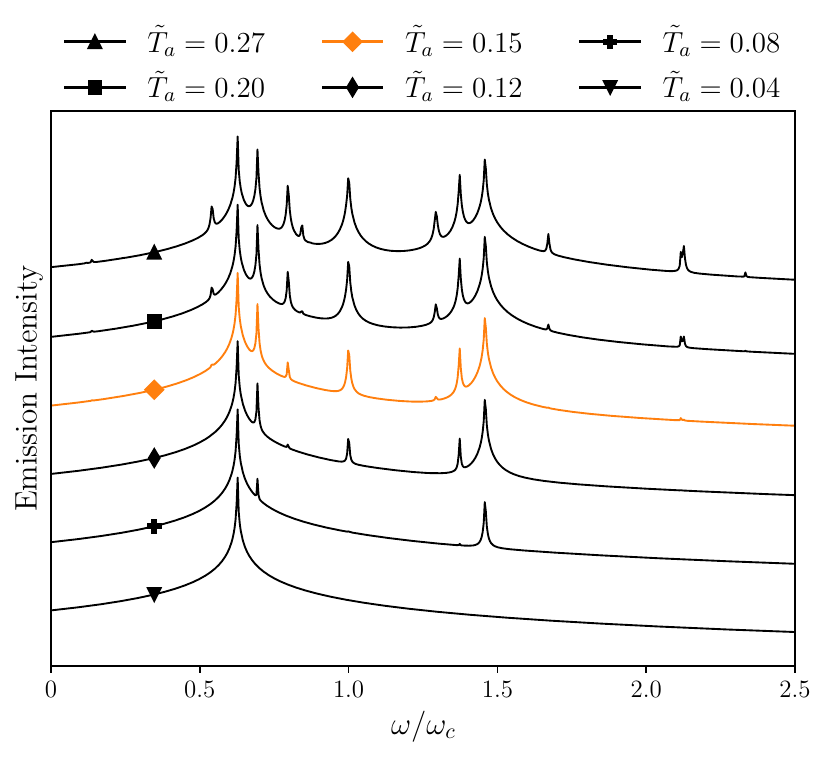}
    \caption{
    Emission spectra for a system with $N=2$ quantum emitters at a fixed normalized coupling strength $\lambda = 0.3$.
    The emission spectra are calculated at different effective temperatures, shown in the legend on top of the figure.
    The emission spectra are vertically shifted along the emission intensity axis, which is in a logarithmic scale.
    The orange spectrum line corresponds to the temperature $\tilde{T}_a = 0.15$, which is used for all the other emission spectra.
    }
    \label{fig:diff_temperature}
\end{figure}
\begin{figure*}
    \centering
    \includegraphics[width=0.9\linewidth]{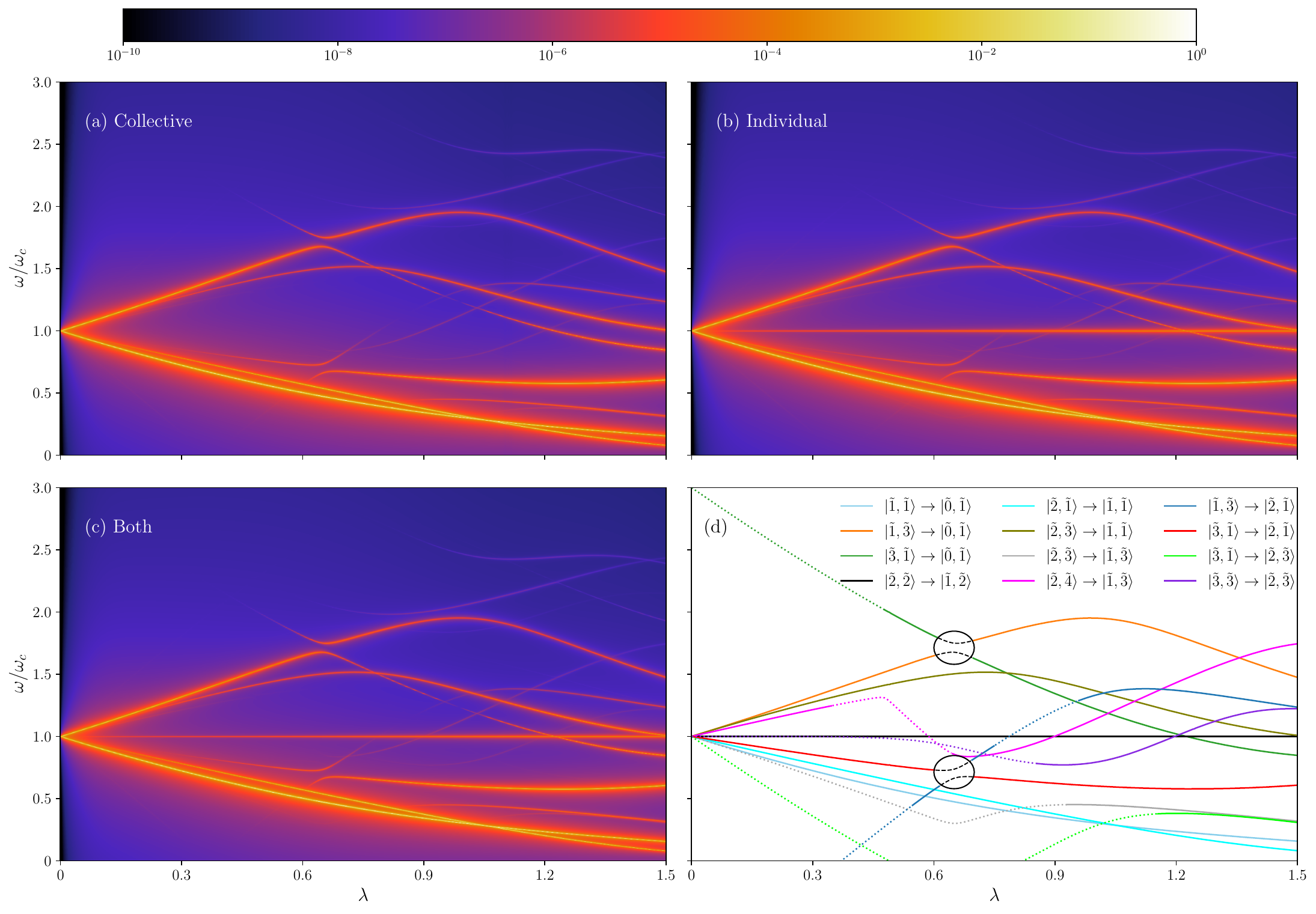}
    \caption{
    (a-c) Emission spectra for a system of $N=2$ quantum emitters as a function of the normalized coupling strength.
    The light-matter system interact with the reservoir in different configurations: (a) collective dissipation channel for the TLSs, (b) individual channels for each TLS, (c) a configuration where both are present.
    (d) Lowest level transitions' energies of the system as function of $\lambda$.
    The continuous curves represent the transition energies observable in the spectra (panels a-c), whereas the dotted curves correspond to transition lines that exhibit weak emission or are optically inactive.
}
    \label{fig:2atom_spectra}
\end{figure*}
\begin{figure}
    \centering
    \includegraphics[width=1\linewidth]{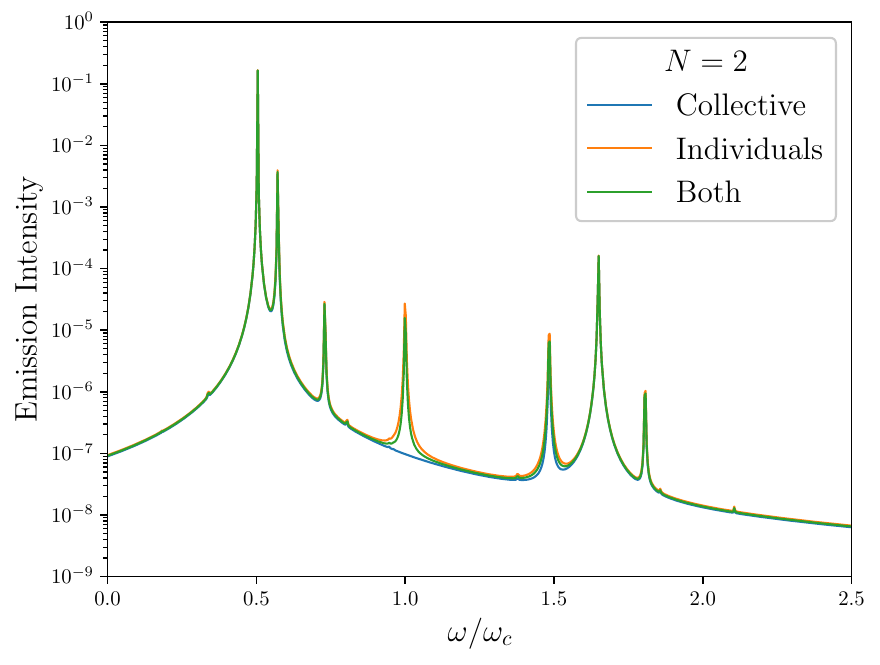}
    \caption{
    Comparison between the emission spectra calculated considering a collective (blue line), individual (orange line), and a combination of the two (green lines) dissipation channels for a system of $N=2$ quantum emitters, for $\lambda=0.6$.
    }
    \label{fig:spettro2D_2a_g06}
\end{figure}

\begin{figure*}
    \centering
    \includegraphics[width=1\linewidth]{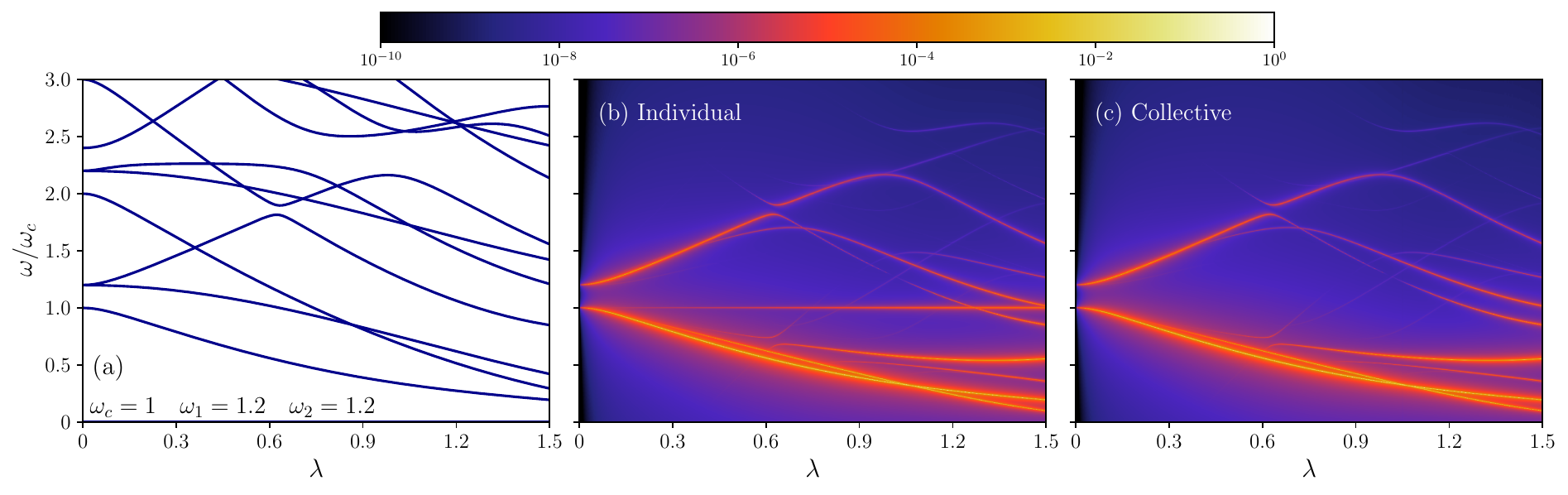}
    \caption{
    (a) Lowest energy eigenvalues of the generalized Dicke Hamiltonian for a system of $N=2$ TLSs detuned with respect to the cavity frequency: $\omega_{1,2} = 1.2\,\omega_c$.
    (b-c) Emission spectra obtained considering individual (b) and collective (c) dissipation channels for the emitters.
    Both eigenvalues and emission spectra are obtained as a function of the coupling strength $\lambda$.
    }
    \label{fig:detuning_cavity}
\end{figure*}

\begin{figure*}
\centering
    \includegraphics[width=0.75\linewidth]{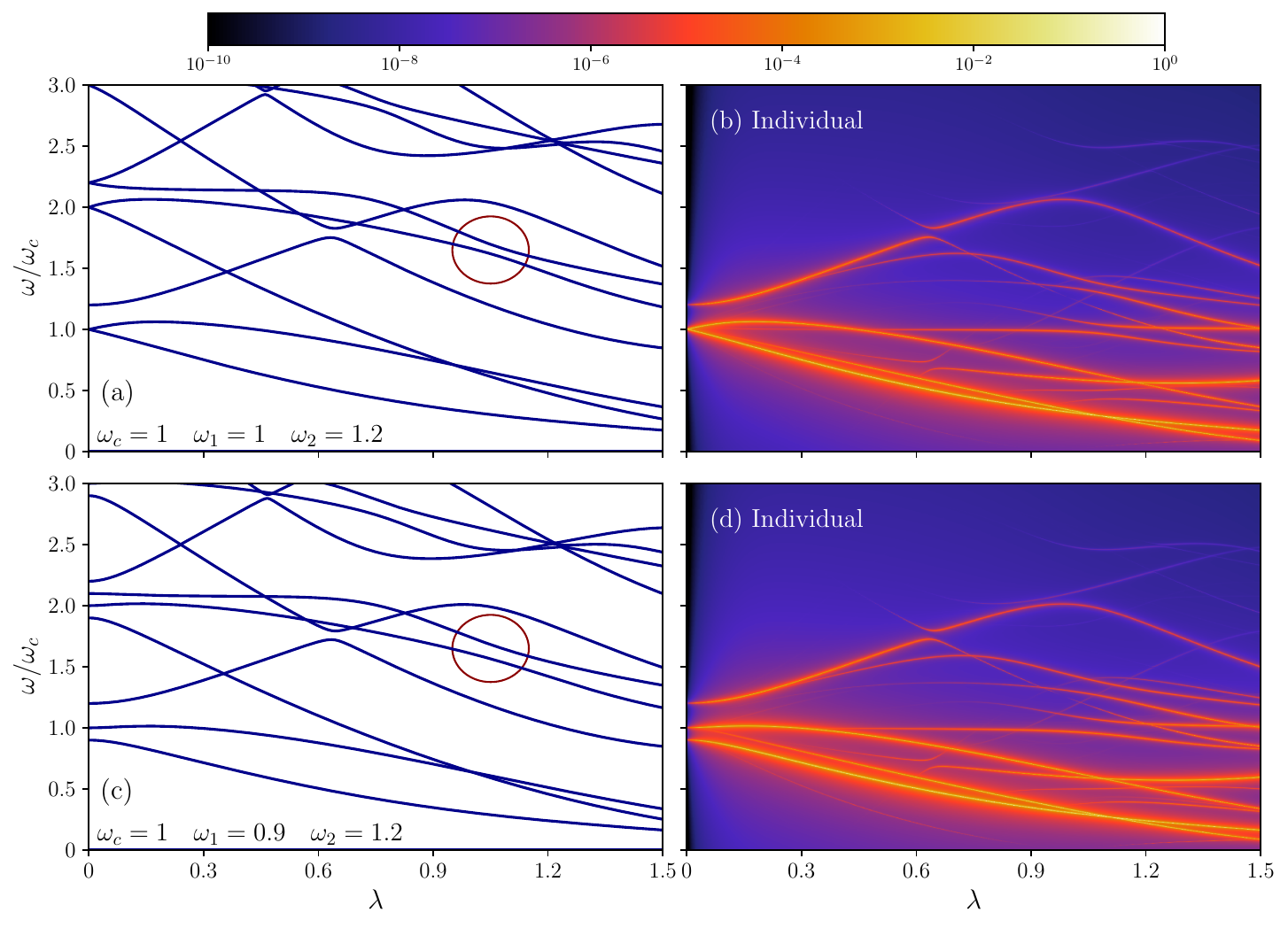}
        \caption{ 
        Energy and emission spectra for $N=2$ TLSs with different transition frequencies as a function of $\lambda$. Parameters: (a,b) $\omega_1 =\omega_{c}$ and $\omega_2 = 1.2\,\omega_{c}$; (c-d) $\omega_1 = 0.9\,\omega_c$, $\omega_2 = 1.2\,\omega_c$.
        }
    \label{fig:detuning_full}
\end{figure*}
%

%
\subsection{Two Emitters} \label{subsec:two_spectra}

In this subsection, we study the two-emitter system  \cite{mercurio2022regimes, akbari2023generalized, jaako2016ultrastrong}.
\figuref{fig:diff_temperature} presents emission spectra calculated at fixed normalized coupling strength ($\lambda=0.3$) for various effective temperatures ranging from $\tilde T_a=0.04$ to $\tilde T_a=0.27$.   
The orange line corresponds to the particular temperature $\tilde T_a = 0.15$ used for the calculation of all the subsequent emission spectra presented.
This temperature, though somewhat arbitrary, serves as a useful compromise: it is high enough to prevent an almost exclusive population of the ground state, yet not so high to excite too many higher energy states.
The  peaks in \figref{fig:diff_temperature} occur at the allowed transition frequencies of the system, which will be more clearly illustrated later in the discussion of \figref{fig:2atom_spectra}(d). As expected, increasing the temperature, enables the population of higher energy levels, giving rise to additional emission lines.

\figuref{fig:2atom_spectra} presents the emission spectra as a function of the normalized coupling strength.
Three cases are considered, based on the specific dissipation channels of the matter subsystem: collective, individual, and a combination of both. This system exhibits a significantly different behavior compared to the single-emitter case. 
In particular, as discussed below, increasing the number of emitters leads to more energy eigenstates,
which give rise to additional emission lines.
While the three cases in \figref{fig:2atom_spectra} largely exhibit similar emission features, a key difference is  evident.
In particular, an additional constant emission line at the frequency $\omega=\omega_c$  in the presence of individual-dissipation reservoirs [\figref{fig:2atom_spectra} \,(b,c)] is visible already at rather small interaction strengths $\lambda$. This constant emission line originates from transitions between states involving antisymmetric states of the matter subsystem. It turns out that these transitions are forbidden for systems interacting with the bath collectively.
This behavior can better understood by comparing the spectra with the lowest energy transition energies (as a function of $\lambda$) shown in \figuref{fig:2atom_spectra}\,(d).
Continuous curves describe the transition energies which can be observed in the spectra, while the dotted curves correspond to the dark or poorly emitting transition lines. 


The most relevant transitions observable in the emission spectra in \figuref{fig:2atom_spectra} involve the ground state and the states with excitation numbers up to $n=3$. 
In particular, three of these transitions occur towards the ground state  $|\tilde{0},\tilde{1}\rangle$, which are: $|\tilde{1},\tilde{1}\rangle \rightarrow |\tilde{0},\tilde{1}\rangle$, $|\tilde{1},\tilde{3}\rangle \rightarrow |\tilde{0},\tilde{1}\rangle$, and $|\tilde{3},\tilde{1}\rangle \rightarrow |\tilde{0},\tilde{1}\rangle$. They represent some of the most intense emissions in the spectra along with those originating from higher excited levels that decay to the level $\Tket{1}{1}$.
As already stated, the emission line at frequency $\omega = \omega_c$ [see \figuref{fig:2atom_spectra}(b,c)] , visible only in the presence of individual qubit reservoirs, originates from the transitions between states involving antisymmetric states of the matter subsystem, i.e. $\{\Tket{2}{2}\rightarrow\Tket{1}{2}, \Tket{3}{2}\rightarrow\Tket{2}{2}\dots |\tilde{c}\!+\!1,\tilde{2}\rangle\rightarrow\Tket{c}{2}\}$. In particular, at low temperature, the main contribution is given by the first transition, $\Tket{2}{2}\rightarrow\Tket{1}{2}$. 
The appearance of this emission line is determined by the individual interaction of the quantum emitters with their baths, which enables the population of such (otherwise dark) states, as well as their radiative decay into one another.
These are $j=0$ states, which are present only in systems with an even number of quantum emitters, as only systems with integer total angular momentum  can yield a zero eigenvalue. In the $N=2$ case, these are antisymmetric states with $n$ photons, thus given by
\begin{equation}
    \ket{n,j=0,m=0}  = \ket{n} \otimes \frac{1}{\sqrt{2}}( \ket{e,g} - \ket{g,e} )  \, .
\end{equation}
They correspond to the states $| \tilde c\!=\!n\!+\!1, \tilde 2 \rangle$, with $\tilde{c} \geq N/2 \in \mathbb{N}$.
Since these states are also eigenstates of the interaction Hamiltonian with zero eigenvalue, they are not modified by the light-matter interaction, giving rise to energy levels which are independent on $\lambda$.
Further evidence is shown in \figref{fig:Eigenspectrum_few_all}, which displays the energy levels for systems with up to six emitters. Specifically, \figref{fig:Eigenspectrum_few_all}(b, d, f) shows the presence of equally-spaced energy levels starting at $\omega=n\omega_c$ (for $\lambda=0$), with $n \in \mathbb N$, 
which, as a function of $\lambda$, share a common curvature, determined uniquely by the energy of the ground state (used as a reference).
Transitions between these levels give rise to the emission line at $\omega=\omega_c$. 
In contrast, the case with an odd number of emitters in \figurefs{fig:Eigenspectrum_few_all}(a, c, e), do not show this behavior.
A general scenario, shown in \figref{fig:2atom_spectra}(c), consists of quantum emitters which can dissipate both individually and collectively. The resulting spectra are quite similar to those observed in the individual dissipation case.

In \figref{fig:spettro2D_2a_g06}, we present a cut at $\lambda=0.6$ of the emission spectra, which further highlights the presence of a rather intense peak at the cavity frequency, $\omega=\omega_c$, when individual reservoirs are considered. On the other hand, the other peaks exhibit similar intensities.
However, upon closer inspection for specific coupling strengths we observe that some peaks in the combined dissipation case are slightly less intense compared to the individual reservoirs case.  
They correspond to peaks where the difference between the collective and individual dissipation cases is greater. 
This observation suggests that when each emitter has its own non-negligible individual dissipation channel, the presence of a collective reservoir has only a minimal impact on the system’s overall emission.

%
%

So far, we have calculated emission spectra at zero detuning ($\omega_1 = \omega_2= \omega_c$ for all the qubits). We will now analyze the consequences of detuning in such systems.
Considering a first case where $\omega_1 =\omega_2 \neq \omega_c$, we observe the partial removal of the degeneracy at $\lambda=0$ due to the emitters-cavity detuning. 
The energy levels corresponding to the dark states have eigenfrequencies $\omega_1 + k \omega_c$, with $k \in \mathbb{N}$. Given their harmonicity with frequency $\omega_c$, the emission line at $\omega = \omega_c$ is still present in the spectra (see \figref{fig:detuning_cavity}).

\begin{figure*}[t]
    \centering
    \includegraphics[width=0.91\linewidth]{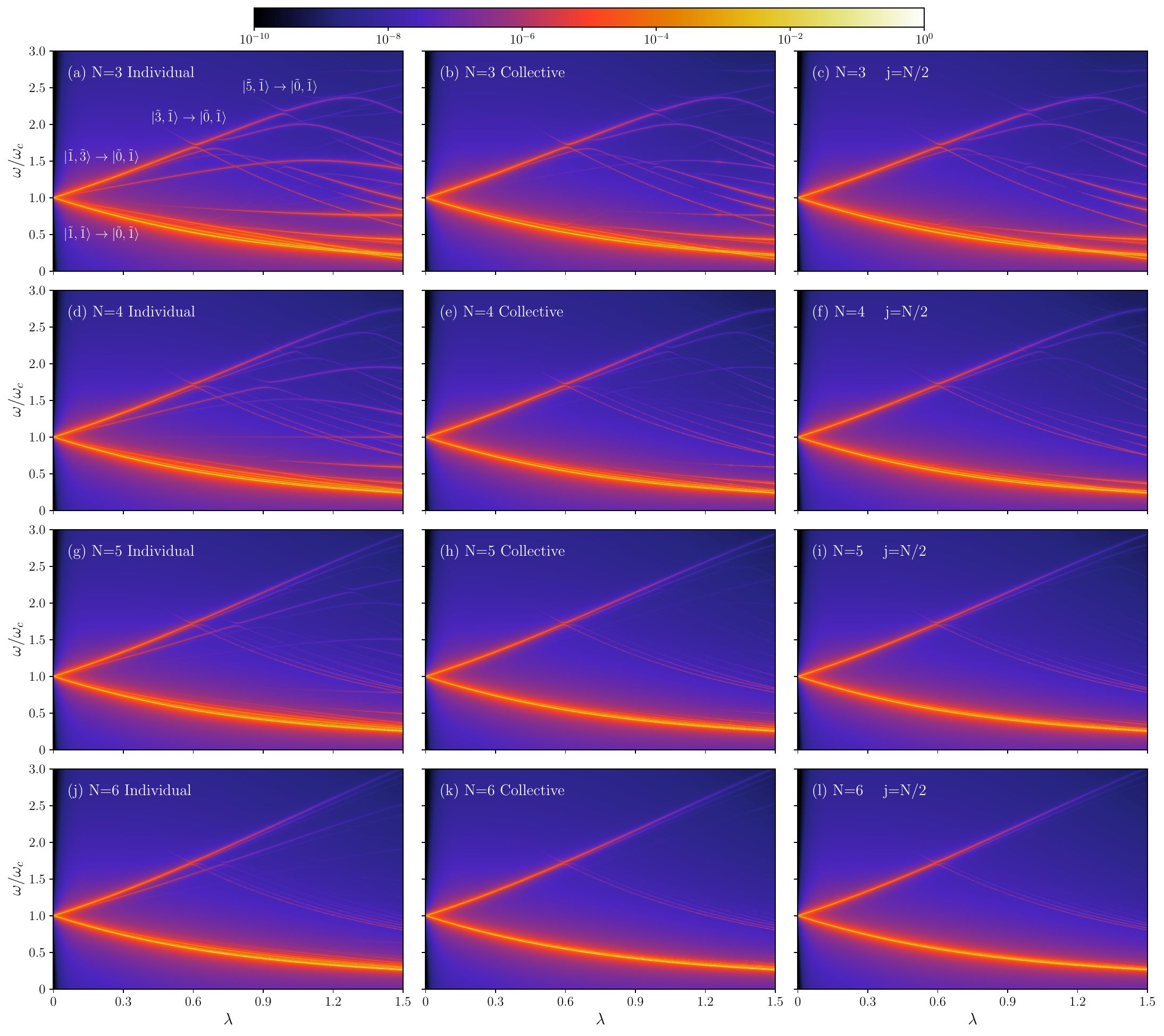}
    \caption{
    Emission spectra for systems composed of $N=3\,$-$\,6$ as a function of the normalized coupling strength. Panels (a,d,g,j) display emission spectra obtained considering individual reservoirs for the TLSs; panels (b,e,h,k) consider a single collective reservoir; panels (c,f,i,l) display collective reservoirs spectra with transitions only between states of maximum total angular momentum.
    }
    \label{fig:Spectra_3_6_full}
\end{figure*}

A second interesting case arises the two emitters are no longer identical, specifically when one is detuned from the other, i.e. $\omega_1 = \omega_c \neq \omega_2$.  In what follows, we consider $\omega_2 = 1.2\, \omega_c$.
\figuref{fig:detuning_full}(a) presents the eigenvalue spectrum as a function of the coupling strength. Notably, the dark states are absent, due to the inability to form an antisymmetric state. Indeed, the frequency mismatch prevents the construction of a collective angular momentum operator, as the emitters are no longer indistinguishable. Consequently, new hybridized states emerge, evidenced by an avoided level crossing around $\lambda \approx 1.08$.
\figuref{fig:detuning_full}(b) displays the emission spectra with individual dissipation channels, since collective decay operators no longer exist in non-degenerate cases. As a result, the previously observed constant emission line at $\omega = \omega_c$ is absent, and instead follows the avoided crossing highlighted by the system's eigenstates.

A final scenario considers the case where all subsystems are detuned. Specifically, we set $\omega_c = 1$, $\omega_1 = 0.9\,\omega_c$, and $\omega_2 = 1.2\,\omega_c$. In \figref{fig:detuning_full}(c), we observe that, at $\lambda = 0$, the eigenvalues within the same excitation manifold, $\tilde{c}$, are already split due to the lack of resonance among the subsystems.
The corresponding emission spectra, shown in \figref{fig:detuning_full}(d), reflect a combination of features discussed in the previous two cases, contrasting clearly with the resonant scenario presented in \figref{fig:2atom_spectra}. In particular, while an avoided level crossing is still present around $\lambda \approx 1.08$, it is less evident in this case, given the increased detuning between the emitters.

\subsection{Few Emitters} \label{subsec:few_spectra}

As the number of emitters increases, the number of energy levels within a given spectral region also grows, adding complexity and making the system more challenging to analyze. However, at the effective temperature considered $\tilde T_a = 0.15$, many transitions are not sufficiently populated. As a result, the emission spectra tend to simplify, with a reduction in the number of prominent emission lines. As we will see, this behavior can be understood by noting that, for increasing $N$, the most active transitions in the system begin to resemble those of a harmonic system (see subsequent subsections).

\figuref{fig:Spectra_3_6_full} displays the emission spectra for systems composed of $N=3\,$-$\,6$ coupled to a single cavity mode. As in the previous subsection, we consider these systems either with only individual dissipation channels (\figref{fig:Spectra_3_6_full}(a,d,g,j)) or collective reservoirs (\figref{fig:Spectra_3_6_full}(b,e,h,k)).
As the number of emitters increases, many emission lines become significantly less intense. In contrast, the transitions $\Tket{1}{1}\rightarrow\Tket{0}{1}$ and $\Tket{1}{3}\rightarrow\Tket{0}{1}$ continue to dominate as the brightest emission lines. Indeed, these lines are associated to the transitions between eigenstates belonging to the manifold of maximum angular momentum $j=N/2$.
We notice the presence of several avoided level crossings along the emission line $\Tket{1}{3}\rightarrow\Tket{0}{1}$, due to the interaction of states of same parity with different excitation number. While number of avoided crossings increases as $N$ becomes larger, they become progressively narrower, scaling as $N^{-1/2}$, to the point where they are barely distinguishable, as evidenced in panels (g,h).

\begin{figure}[t]
    \centering
    \includegraphics[width=1\linewidth]{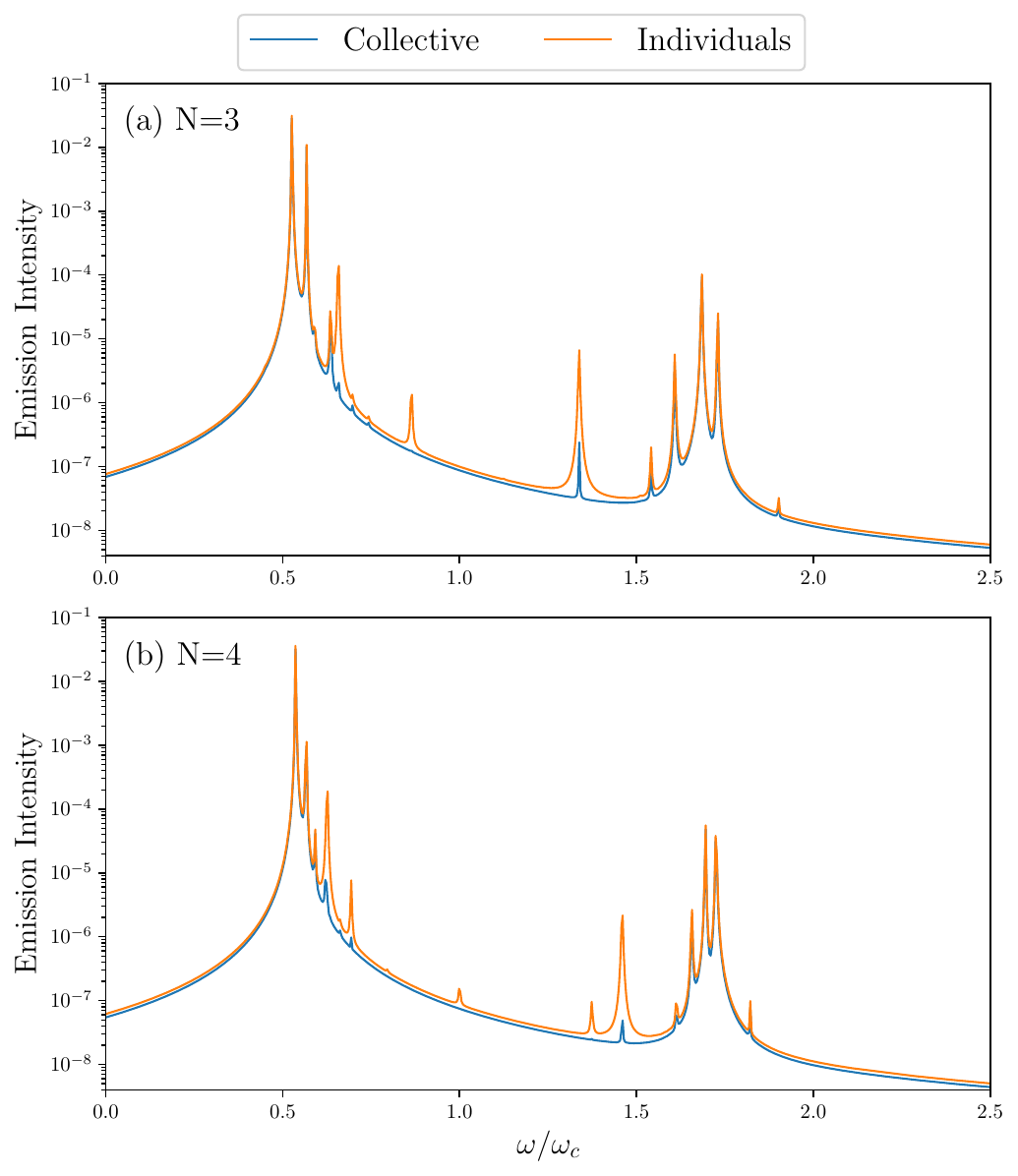}
    \caption{
    Comparison between the emission spectra calculated considering a collective (blue line) and  individual (orange line) dissipation channels for the $N=3$ emitters (a) and for $N=4$ emitters (b) at  $\lambda=0.6$.
    }
    \label{fig:spettro_3e4}
\end{figure}

\figuref{fig:spettro_3e4} presents the emission spectra for $N = 3$ (panel a) and $N = 4$ (panel b) emitters at $\lambda = 0.6$, comparing the system's behavior when coupled to collective (blue) and individual (orange) reservoirs. Overall, the spectra reveal that collective reservoirs produce narrower peaks, and certain transitions become active only at higher coupling strengths compared to individual baths.
The most notable distinction, however, is the presence of the emission peak at $\omega = \omega_c$ (at any coupling strength) for systems with an even number of emitters. This feature stems from transitions between dark states, as discussed previously. Nevertheless, the intensity of this emission line diminishes with increasing $N$, since the involved dark states lie higher in energy and are therefore less thermally populated under weak pumping conditions.

As a final note, we analyze the emission spectra by considering only the levels within the manifold of maximum total angular momentum, $j = N/2$, which includes the ground and first excited states and exhibits the strongest coupling to the photonic mode \cite{Emary2003}. This restriction of the matter Hilbert space becomes computationally advantageous when dealing with systems composed of many emitters, as explored in the following subsection, while still preserving all the relevant physical features. 
In \figref{fig:Spectra_3_6_full}(c,f,i,l), we present the emission spectra obtained by restricting the matter Hilbert space to the $j = N/2$ manifold, for systems with $N = 3\,$-$\,6$ emitters coupled to collective reservoirs. 
We observe that, as $N$ increases, the differences between the full and restricted spectra become irrelevant. This is because the emission lines associated with $j < N/2$ manifolds become negligible in the limit $N \to \infty$. Based on these observations, we restrict our analysis to the $j = N/2$ manifold in the subsequent subsection.

\subsection{Many Emitters and Thermodynamic Limit} \label{subsec:Many_spectra}

\begin{figure*}
    \centering
    \includegraphics[width=0.75\linewidth]{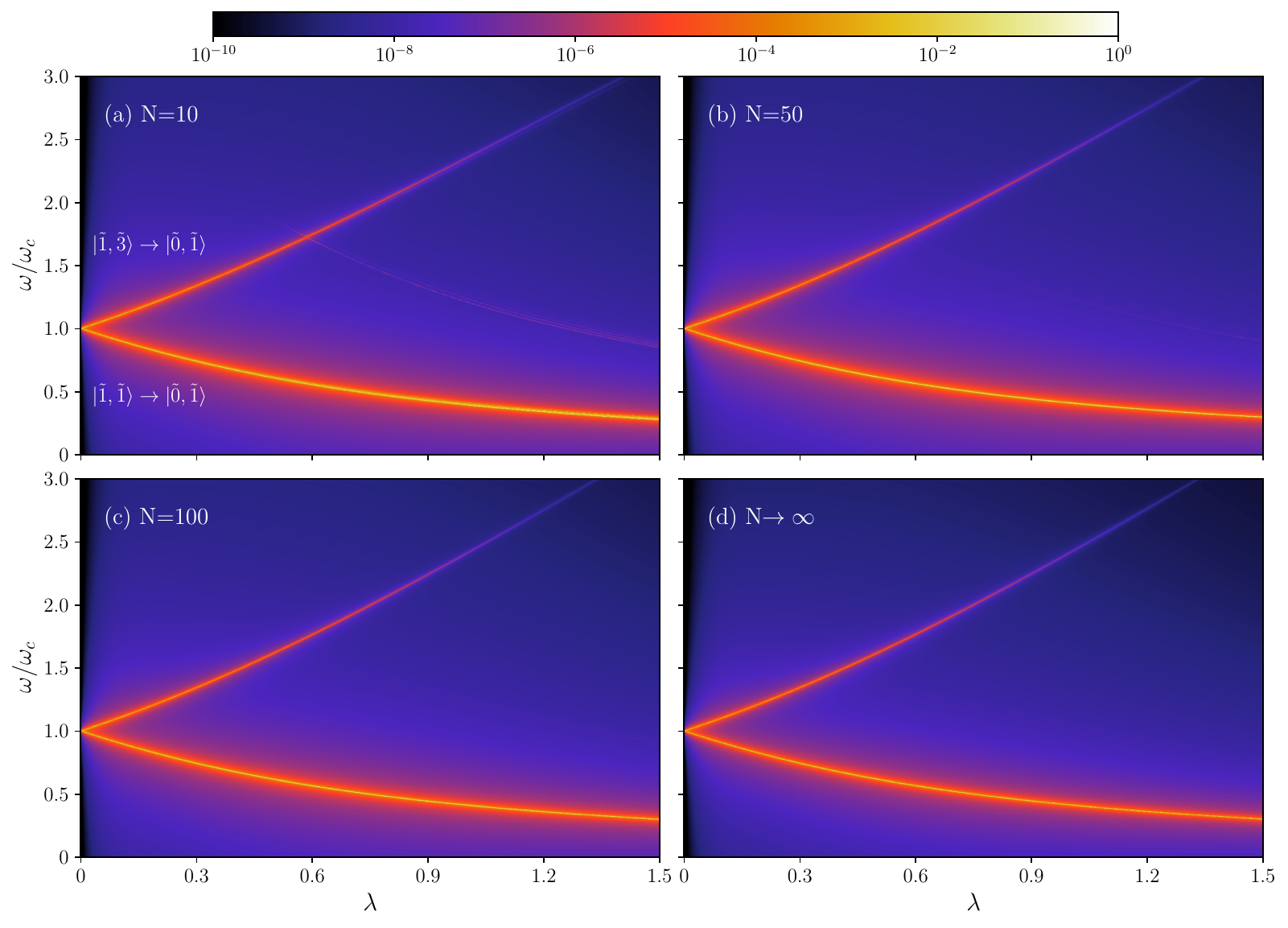}
    \caption{
    Emission spectra for systems of $N=10$ (a), $N=50$ (b), $N=100$ (c) and $N\rightarrow\infty$ (d) TLSs as function of the normalized coupling strength $\lambda$, considering a collective dissipation channel.
    }
    \label{fig:SP_ManyHop}
\end{figure*}

In this subsection, we examine the emission spectra of systems composed of a large number of emitters ($N = 10, 50, 100$), approaching the thermodynamic limit $N \to \infty$, whose energy levels were shown in \figref{fig:Eigenspectrum_MANY}. As previously mentioned, the analysis is restricted to the maximum angular momentum manifold $j = N/2$.
As illustrated in \figuref{fig:SP_ManyHop}, the number of visible emission lines decreases with increasing $N$, ultimately leaving only the two most intense emission lines, corresponding to the transitions $\Tket{1}{3} \rightarrow \Tket{0}{1}$ and $\Tket{1}{1} \rightarrow \Tket{0}{1}$. Furthermore, the avoided level crossings, still faintly visible for $N = 10$ (see \figuref{fig:SP_ManyHop}(a)), vanish entirely for larger $N$.
The overall behavior displayed in these spectra closely resembles that of the lower and upper polariton modes in the thermodynamic limit (see \figuref{fig:SP_ManyHop}(d)). Indeed, as derived in \eqref{eq:Hhop_C} and \eqref{eq:Hhop_D}, the collective dynamics of the quantum emitter ensemble can be effectively described through bosonization, leading to the bosonic Hopfield model. The disappearance of higher-order transitions is a clear sign of the system's tendency toward harmonicity in the thermodynamic limit.
These conclusions hold within the weak excitation regime considered here. At higher effective temperatures, where more excited states become significantly populated, deviations from the Hopfield model may reemerge, and a larger number of emitters would be required to maintain the same level of agreement.

%
%

\section{CONCLUSIONS} \label{thend}

In this work, we have presented a comprehensive analysis of systems composed of 
$N$ two-level quantum emitters coupled to a single cavity mode for light-matter coupling strengths ranging from the weak coupling to the ultrastrong and deep-strong coupling regimes. In particular, we have systematically explored the characteristics of the energy spectrum and as a function of both the coupling strength and the number of emitters. Furthermore, by employing a gauge-invariant theoretical framework suitable for open quantum systems and valid for arbitrary coupling strengths, we have systematically explored the emission properties under incoherent excitation of the quantum emitters, extending our analysis also to systems with different numbers of quantum emitters.
We have also investigated the impact of the nature of the coupling between the system and the external reservoir on the emission spectra , i.e. considering either a local (individual) reservoir for each of the emitters, a collective reservoir for all them, or a co-existence of both. In particular, we found that, in the presence of individual reservoirs, when all the emitters have the same transition frequency, a sharp emission line at the resonant frequency of the resonator can be observed for an even number of emitters. At low effective temperatures, this effect is particularly pronounced for $N=2$, while at higher temperatures it becomes appreciable even for $N \ge 4$.

Our analysis also shows that, as the number of emitters increases, the system undergoes a smooth transition towards an effective model of two interacting bosonic modes, recovering in the end the well-known Hopfield model.
More specifically the system undergoes a qualitative transformation: from discrete, anharmonic, atom-like spectral features to a regime where the ensemble's relevant excitations behave as an effective bosonic field. This emergent bosonic character provides a bridge between microscopic and macroscopic regimes of cavity QED. 

The results here presented can be useful for the analysis of experiments in which the USC regime is achieved with a few quantum emitters coupled to plasmonic nanoresonators.
Moreover, these findings provide new insights into the interplay between collective effects, dissipation, and ultrastrong light-matter coupling, offering a unified perspective that connects few-body quantum optics with many-body quantum electrodynamics. 
This work could prove valuable for the creation of quantum technologies, particularly those based on superconducting qubits and collective light-matter interactions.


\appendix

\section{The $\lambda \to \infty$ limit: Perturbation of the Displaced Harmonic Oscillator} \label{app:harmonic}

In this appendix, we show that the eigenstates of ${\cal \hat H}_{0}'$ [see \eqref{DSC}] can be obtained by applying the displacement operator $\hat D(\hat\alpha)$ to
the states $\ket{n,j,m}$, as
\begin{equation} \label{eq:appA_ntilde_Dn}
   \ket{\tilde{n}_m,j,m} = \hat D^\dagger(\hat\alpha) \ket{n,j, m}\, ,
\end{equation}
where $n$ is the (undisplaced) photon number ($\hat a^\dagger \hat a |n \rangle = n  |n \rangle$), $j$ is the total angular momentum quantum number and $|m \rangle$ is an eigenstate of $\hat J_x$: $ \hat J_x |m \rangle = m |m \rangle$.
Here, the operator $\hat\alpha = 2i\lambda\hat J_x$, acting on the state $\ket{n,j,m}$, assumes the value of the corresponding eigenvalue $2i\lambda m$, so, \eqref{eq:appA_ntilde_Dn} can be rewritten as follows
\begin{equation}\label{dim}
\ket{\tilde{n}_m,j,m} = \hat D^\dagger(2i\lambda m) \ket{n,j, m}\, .
\end{equation}
To derive this result, we extend the definition of the displacement operator $\hat D(\alpha)$, which is a function  of the complex variable $\alpha$,
by promoting this c-number  to an operator: $\hat\alpha = 2i\lambda\hat J_x$.
In this framework,  the destruction operator of the displaced harmonic oscillator  $\hat{\cal A} = \hat a + \hat\alpha$,  can be obtained from the transformation
\begin{equation}
    \hat D^\dagger(\hat\alpha) \opa \hat D(\hat\alpha) = \opa + \hat\alpha = \hat{\cal A} \, ,
\end{equation}
where the displacement super-operator is explicitly given by
\begin{equation}
\begin{split}
    & \hat D(\hat\alpha) = e^{\hat\alpha \opad - \hat\alpha^\dag \opa} = e^{\hat\alpha (\opad + \opa)} \, , \\
    & \hat D(\hat\alpha) = \hat D^\dagger(-\hat\alpha) \, ,
\end{split}
\end{equation}
where we used $\hat\alpha^\dagger = -\hat\alpha$.
The ground state of the displaced system is defined through the action of the displaced destruction operator on the ground state of the Dicke model
\begin{equation} 
    (\opa + \hat\alpha)\ket{\tilde{0}_m,j,m} = \hat{\cal A}\ket{\tilde{0}_m,j,m} =  0
    \, .
\end{equation}
Consequently, it follows that
 \begin{equation} \label{eq:app_a_alpha1}
    \opa \ket{\tilde{0}_m,j,m} = -\hat\alpha\ket{\tilde{0}_m,j,m}\,.
\end{equation}
We know that the destruction operator $\hat a$ acting on a generic coherent state, $|\hat\alpha_m , j,m\rangle \equiv |\alpha_m , j,m\rangle$, gives
\begin{equation} \label{eq:app_a_alpha2}
    \hat a\ket{-\hat\alpha_m,j,m} = -\hat\alpha\ket{-\hat\alpha_m,j,m}\,.
\end{equation}
So, comparing Eq.s (\ref{eq:app_a_alpha1}) and (\ref{eq:app_a_alpha2}), we obtain the expression of the ground state for the displaced harmonic oscillator in terms of a coherent state of the original system
\begin{equation} 
   \ket{\hat 0_m,j,m}= \ket{-\hat\alpha_m,j,m}\,.
\end{equation}
In addition we, obtain
\begin{equation}
    \ket{\tilde{0}_m,j,m}  =\hat D(-\hat\alpha)\ket{0,j,m}= \hat D^\dagger(\hat\alpha)\ket{0,j,m} \,.
\end{equation}

%
Now, we are able to express an arbitrary eigenstate of the displaced harmonic oscillator as, 
%
\begin{eqnarray} \label{eq:app_ntilde_n}
    \ket{\tilde{n}_m,j,m} &=& \frac{(\hat A^\dagger)^{n}}{\sqrt{n!}} \ket{\tilde{0}_m,j,m} = \\
    &=&  \frac{\hat D^\dagger(\hat\alpha) (\opad)^{n}\hat D(\hat\alpha)}{\sqrt{n!}}\hat  D^\dagger(\hat\alpha)\ket{0,j,m} \!= \nonumber \\
    &=& \frac{\hat D^\dagger(\hat\alpha) (\opad)^{n}}{\sqrt{n!}} \ket{0,j,m} 
   = \hat D^\dagger(\hat\alpha)\ket{n,j,m}\,.\nonumber
\end{eqnarray}
Introducing $\hat\alpha = 2i\lambda\hat J_x$ on the above equation, we obtain \eqref {dim}.

Now we show how the dressed photon excitation number $\tilde{n}$  for the displaced state is related to the bare photon excitation label.   
We have
\begin{equation}
\begin{split}
    \tilde{n}\ket{\tilde{n}_m,j,m} &= {\cal A}^\dagger {\cal A} \ket{\tilde{n}_m,j,m} \,,\\
    &= \hat D^\dagger(\hat\alpha) \opad \hat D(\hat\alpha)\hat D^\dagger(\hat\alpha) \opa \hat D(\hat\alpha)\ket{\tilde{n}_m,j,m} \,,\\
    &= \hat D^\dagger(\hat\alpha) \opad\opa \hat D(\hat\alpha)\ket{\tilde{n}_m,j,m} \,.
\end{split}
\end{equation}
Hence, multiplying $\hat D(\hat\alpha)$ on the left, to both sides
\begin{equation}\label{nuova}
  \tilde{n}\hat D(\hat\alpha)\ket{\tilde{n}_m,j,m} = \opad\opa \hat D(\hat\alpha)\ket{\tilde{n}_m,j,m}  \,.
\end{equation}
From the relation obtained in \eqref{eq:app_ntilde_n}, we have 
\begin{eqnarray} \nonumber
  \ket{n,j,m}=\hat D(\hat\alpha) \ket{\tilde{n}_m,j,m} \, .\nonumber
\end{eqnarray}
We finally obtain from \eqref{nuova}
$$ \opad\opa\ket{n,j,m} = \tilde{n}\ket{n,j,m}\,.$$
This clearly shows that we have $\tilde{n} = n$.

When $\lambda >1$, but we are not in the limit $\lambda \to \infty$, it is possible to derive analytically approximate eigenvalues and eigenstates of the generalized Dicke [\eqref{eq:H_D_Dicke}] by employing perturbation theory. 
In this limit, the Hamiltonian can be regarded as that of a perturbed displaced harmonic oscillator where the atomic bare energy term $\hbar \omega_a \hat{J}_z$, can be treated as a small perturbation of the system. 
 We have, at resonance ($\omega_c=\omega_a$)
  $  {\cal \hat H}_{\rm D} = \hbar \omega_c \hat A^\dag \hat A + \hbar \omega_a \hat{J}_z$.
Due to the degeneracy of the eigenstates of $\hat{\cal H}_0' =\hbar \omega_c \hat A^\dag \hat A$, we use the degenerate perturbation theory.
Specifically, we need to calculate the mean values of the $\hbar\omega_a\hat{J}_z$  for the eigenstates $\ket{\tilde n_m,j, m}$ of the unperturbed Hamiltonian $\hat{\cal H}_0'$ [see also \eqref{DSC}]. We obtain
\begin{eqnarray} \label{eq:appA1}
   && \bra{\tilde n_m,j,m} \hat{J}_z \ket{\tilde n_{m'},j',m'} 
     =
    \bra{\tilde n_m,j,m} \hat{J}_z \ket{\tilde n_{m'},j,m'} \delta_{j,j'}\nonumber \\ 
    &&= \bra{n,j,m}\hat D^\dagger(\hat\alpha) \hat{J}_z  \hat D(\hat\alpha)\ket{n,j,m'} \,,
\end{eqnarray}
where we used  \eqref{eq:appA_ntilde_Dn}. 
The operator in the r.h.s. of \eqref{eq:appA1}, $\hat D^\dagger(\hat\alpha) \hbar\omega_a\hat{J}_z \hat D(\hat\alpha)$, can be expanded using the Baker-Campbell-Hausdorff formula
\begin{equation} \label{eq:pert_expa1}
\begin{split}
    &
    \hat D^\dagger(\hat\alpha) \hat{J}_z \hat D(\hat\alpha) = 
    \{ \hat J_z + 2i\lambda(\opad + \opa)[\hat J_x,\hat J_z] + \\
    &\frac{(2i\lambda(\opad + \opa))^2}{2!}[\hat J_x,[\hat J_x,\hat J_z]]  + \dots + \\& \frac{(2i\lambda(\opad + \opa))^k}{k!}[\underbrace{\hat J_x,[\dots,[\hat J_x}_{k-times},\hat J_z] \} = \\
    & =  
    \{ \hat J_z + 2i\lambda(\opad + \opa)(-i \hat J_y) + \frac{(2i\lambda(\opad + \opa))^2}{2!} \hat J_z+\\& + \frac{(2i\lambda(\opad + \opa))^3}{3!} (-i \hat J_y) +\dots \}=\\
    & =  
    \{ \hat J_z \left( 1-\frac{(2\lambda(\opad + \opa))^2}{2!} +\dots\right)
           \\&+\hat J_y\left( 2\lambda(\opad + \opa) -\frac{(2\lambda(\opad + \opa))^3}{3!}\dots\right) \}=
        \\
    & = 
    \hat J_z {\rm cos}[2\lambda(\opad + \opa)] + \hat J_y {\rm sin}[2\lambda(\opad + \opa)] 
    \,.
\end{split}
\end{equation}
%
Here, the cosine and sine operator functions can be expanded in terms of their Taylor series. 
Upon expansion, one encounters terms of the form $(\opad+\opa)^k$, whose explicit evaluation becomes increasingly non-trivial for large values of $k$  due to the non-commutative nature of the creation and annihilation operators. 
The resulting combinations of creation and annihilation operators are subsequently reordered into normal order \cite{chatterjee2023general}:
\begin{equation} \label{eq:cosine_expansion}
    \begin{split}
        & {\rm cos}(2\lambda(\opad + \opa))\ket{n} = \sum_{k=0}^\infty \frac{(-1)^k[2\lambda(\opad+\opa)]^{2k}}{(2k)!}\ket{n} = \\
        & = \sum_{k=0}^\infty \frac{(-1)^k(2\lambda)^{2k}}{(2k)!} \sum_{m=0}^{[2k/2]} \frac{(2k)!}{m! 2^m (2k-2m)!} \cdot \\
        & \quad\quad \quad \cdot \!\sum_{r=0}^{2k-2m} \frac{(2k-2m)!}{r!(2k-2m-r)!}(\opad)^{(2k-2m-r)} \opa^r \ket{n} \,.
    \end{split}
\end{equation}
In a similar way, for the sine operator function, we obtain
\begin{equation}
    \begin{split}
        & {\rm sin}(2\lambda(\opad + \opa))\ket{n} = \sum_{k=0}^\infty \frac{(-1)^k[2\lambda(\opad+\opa)]^{2k+1}}{(2k+1)!}\ket{n} = \\
        & = \sum_{k=0}^\infty \frac{(-1)^k(2\lambda)^{2k+1}}{(2k+1)!} \sum_{m=0}^{[(2k+1)/2]} \frac{(2k+1)!}{m! 2^m (2k+1-2m)!} \cdot \\
        & \,\,\,\, \cdot \!\!\!\!\sum_{r=0}^{2k+1-2m} \frac{(2k+1-2m)!}{r!(2k+1-2m-r)!}(\opad)^{(2k+1-2m-r)} \opa^r \ket{n} \,.
    \end{split}
\end{equation}
Notice that $$ \bra{n,j,m}\hat J_y \sin [2\lambda(\opad + \opa)]\,\ket{n,j,m'}=0 \,, $$ because the creation and annihilation operators appear always in unequal number, i.e., leading to terms of the form  $\bra{n}{\hat a}^{\dag \, k} {\hat a}^{k'}\ket{n}$ with $k\neq k'$, whose value is zero.
In this way, \eqref{eq:pert_expa1} can be reduced to
\begin{equation} \label{eq:appA3}
\begin{split}
   & \bra{n,j,m}\hat D^\dagger(\hat\alpha) \hat{J}_z \hat D(\hat\alpha)\ket{n,j,m'} 
    \\
    & = 
    \bra{n,j,m}\{\hat J_z {\cos}[2\lambda(\opad + \opa)]\,\}\ket{n,j,m'} \,.
\end{split}
\end{equation}
%
Using \eqref{eq:cosine_expansion}, we can calculate explicitly the action of the operator ${\rm cos}[2\lambda(\opad + \opa)]$ on the state $\ket{n,j,m'}$ in the r.h.s of \eqref{eq:appA3}
\begin{equation} \label{eq:appA4}
\begin{split}
    & \cos{[2\lambda(\opad + \opa)]}\ket{n,j,m'}=\sum_{k=0}^\infty (-1)^k(2\lambda)^{2k}\sum_{m=0}^{[2k/2]}\\
    & \sum_{r=0}^{2k-2m} \frac{\sqrt{n!(n+2k-2m-2r)!}}{m! 2^m r!(2k-2m-r)!(n-r)!} \ket{n+2k-2m-2r} \,.
\end{split}
\end{equation}
Note that in the r.h.s of \eqref{eq:appA3}, being the operator $J_z$ acting only on $\ket{m'}$, and using  the result in \eqref{eq:appA4}, we have scalar products such $\braket{n}{n+2k-2m-2r} = \delta_{2k-2m-2r,0}$. 
Hence, it is possible the substitution $r = k-m$ that allows the elimination of the sum over $r$ in \eqref{eq:appA4}. For the r.h.s. of \eqref{eq:appA4} we have
\begin{equation} \label{eq:appA5}
    \sum_{k=0}^\infty (-1)^k(2\lambda)^{2k} 
    \!\sum_{m=0}^{[2k/2]} \!\!\frac{n!}{m! \,2^m ((k-m)!)^2(n-(k-m))!} \ket{n} \,.
\end{equation}
Moreover, the two sums respectively on $k$ and $m$ variables can be separated and rewritten using the Cauchy product formula
\begin{equation}
    \sum_{i=0}^\infty \sum_{j=0}^i a_j b_{i-j} = \sum_{i=0}^\infty a_i \cdot \sum_{j=0}^\infty b_j \,.
\end{equation}
Hence, taking:
\begin{equation}
\begin{split}
    & a_i = \frac{(-1)^m (2\lambda)^{2m}}{m!2^m} \,, \\
    & b_j = \frac{(-1)^{k-m}(2\lambda)^{2(k-m)}}{((k-m)!)^2(n-(k-m))!} \,.
\end{split}
\end{equation}
\eqref{eq:appA5} transforms as
\begin{equation} 
\begin{split}
    & n!\,\sum_{k=0}^\infty \frac{(-1)^k(2\lambda)^{2k}}{k!\,2^k} 
    \sum_{l=0}^{\infty} \frac{(-1)^l(2\lambda)^{2l}}{(l!)^2(n-l)!} \ket{n} =\\
    & =  e^{-2\lambda^2}  \sum_{l=0}^{n} \binom{n} {l}\frac{(-1)^l(4\lambda^2)^{l}}{l!} \ket{n}=e^{-2\lambda^2} L_n(4\lambda^2)
    \,,
   \end{split}
\end{equation}
with $l=k-m$.
The result above is obtained considering that the sum over $k$ is equal to $e^{-2\lambda^2}$, while the second sum is limited by the domain $0\leq l \leq n$ and gives the $n$-th Laguerre polynomial.
Finally, observing that \eqref{eq:appA3} gives the correction at first order $\Delta_E^{(1)}$ in the DSC regime, we can explicitly write
\begin{equation}
\begin{split}
    \Delta_E^{(1)} &= \hbar\omega_a \bra{n,j,m}\hat J_z {\rm cos}[2\lambda(\opad + \opa)]\ket{n,j,m'} = \\
& = \hbar\omega_a \,e^{-2\lambda^2}L_n(4\lambda^2) \bra{j,m}\hat J_z \ket{j,m'}    
    \,.
\end{split}
\end{equation}

To determine the corrections to the eigenvalues of the unperturbed Hamiltonian, this matrix must be diagonalized within the subspace defined by fixed values of $n$, and $j$. Such diagonalization procedure also provides the approximated eigenvectors.

This procedure can be followed by second-order perturbation theory, to include corrections arising from the hybridization of the approximated eigenstates with levels outside their subspace. Such correction $\Delta_E^{(2)}$ is given by
\begin{equation}\label{2ord}
    \Delta_E^{(2)} = \hbar\omega_a\sum_{q\neq n} \frac{ \left| \bra{\tilde{q}_m,j,m} \hat{J}_z \ket{\tilde{n}_m,j,m} \right |^2 } {E_{n} - E_{q} } \,.
\end{equation}
\newline

By following a procedure similar to that described above, we  obtain

\begin{equation}
\begin{split}
      \Delta_E^{(2)} & =  \hbar\omega_a \sum_{q \neq n}^\infty \sum_{m,m'}^{2^N}
    \frac{1}{E_n - E_q}  \\  
    & \left\{  \left| \bra{q}\cos{[2\lambda(\opad + \opa)]}\ket{n}\bra{m}_x\hat J_z\ket{m'}_x + \right.\right. \\  & \quad \left.\left.  \bra{q}\sin{[2\lambda(\opad + \opa)]}\ket{n}\bra{m}_x\hat J_y\ket{m'}_x  \right|^2 \right \} \,.
\end{split}
\end{equation}

The so obtained approximated energy eigenvalues, for $\lambda \gtrsim 1$, results to be in good agreement with the exact numerical calculations presented in this paper (not shown).

\section{Generalized Master Equation} \label{sec:GME}
In this Appendix, we develop the master equation formalism to describe the interaction with the thermal reservoirs of a system constituted by $N$ identical two-level systems ultrastrongly coupled to a single resonant mode of an electromagnetic resonator.

For a closed system described by a Hamiltonian $\hat{\cal H}^S$,  the dynamics of the corresponding open system can be described using the density operator formalism \cite{breuer2002theory}. In this framework, the master equation for the open system, determining the dynamics of the reduced density operator for the system degrees of freedom, can be obtained starting from the  von Neumann equation for a closed enlarged system, including the reservoirs degrees of freedom:
\begin{equation} \label{eq:Liou_von}
    \frac{d}{dt}\hat{\rho}_T(t) = -i[\hat{\cal H}, \hat{\rho}_T(t)] \, ,
\end{equation}
where ${\hat \rho}_T(t)$ is the total density operator and $\hat{\cal H}$ is the total Hamiltonian of the enlarged system.
The total Hamiltonian can be divided as
$${\hat{\cal H} = \hat{\cal H}^S + \hat{\cal H}^R + \hat{\cal H}^I \equiv \hat{\cal H}^0 + \hat{\cal H}^I}\,,$$ 
where $\hat{\cal H}^S$ is the system Hamiltonian, $\hat{\cal H}^R$ is the Hamiltonian for the reservoirs, and $\hat{\cal H}^I$ represents the system-reservoirs interaction Hamiltonian. 
In the generalized master equation (GME) approach, the Liouvillian superoperator 
 ${\cal L}_{\rm gme}$  is introduced  to describe the reduced dynamics of the system \cite{Settineri2018dissipation}. 
 By applying the second-order Born approximation and assuming the Markovian limit, the reduced density operator $\hat \rho$, which describe the (now open) system $\cal S$, is obtained by tracing out the reservoirs degrees of freedom: ${\hat \rho} \equiv {\rm Tr_R}\{\hat \rho_T\}$. Thus, starting from the von Neumann equation
\eqref{eq:Liou_von}, the time evolution of  the reduced density operator governed by the generalized master equation can be written as
\begin{equation} 
    \frac{d}{dt}\hat{\rho}(t) = -i[\hat{\cal H}^S, \hat{\rho}]  +{\cal L}_{\rm gme}(t)\hat{\rho}(t) \, .
\end{equation}
To obtain an explicit form of the Liouvillian superoperator ${\cal L}_{\rm gme}$, we begin by expressing the system operators in the dressed basis of $\hat{\cal H}^S$. In this basis, the projectors onto the eigenspaces determined by a given eigenvalue $\epsilon$ are defined as $\hat{\Pi}(\epsilon) \equiv \ketbra{\epsilon}{\epsilon}$.
 Next, we define the system operator $\hat{S}_i$ in terms of its frequency components as follows
\begin{equation}
\begin{split}
    \hat{S}_i(t) 
    & = \sum_{\epsilon' - \epsilon = \omega} \hat{\Pi}(\epsilon) (\hat{s}_i + \hat{s}_i^{\dagger})  \hat{\Pi}(\epsilon') e^{-i \omega t} \\
    & = \sum_{\epsilon' - \epsilon = \omega} \hat{S}_i(\omega)e^{-i \omega t} \, ,
    \end{split}
\end{equation}
These operator, $\hat{s}_i$ and  $\hat{s}^\dag_i$ are the annihilation and creation operators for the $i$-th subsystem (or lowering and raisng operators for the quantum emitters), which mediate the interaction with the reservoirs \cite{Settineri2018dissipation}.

The operator $\hat{S}_i$ can be further separated into its positive- and negative-frequency components. 
The positive-frequency operator $\hat{S}_i^{(+)}(\omega)$ induces transitions from higher- to lower-energy eigenstates, whereas the negative-frequency operator $\hat{S}_i^{(-)}(\omega)$ induces transitions from lower- to higher-energy eigenstates.
In the case of TLSs, the dissipation operators are given by ${\hat s_{\rm ind,\,k} = \sigma_k^-}$ for the $k$-th individual quantum emitter, and by ${\hat s_{\rm coll} = \frac{1}{\sqrt{N}}\sum_k \sigma_k^-}$ when considering a collective dissipation channel.
These operators constitute the building blocks of the dissipative part of the Liouvillian.
The $\text{Liouvillian superoperator }$ can be written as
\begin{widetext}
\begin{equation} \label{eq:Lgme1}
\begin{split}
\mathcal{L}_{\rm gme} \hat \rho = \frac{1}{2}\sum_i\sum_{(\omega,\omega')>0} \bigg\{ &\Gamma_i (\omega') n (\omega', T_i) \mleft[ \hat S^{(-)}_i (\omega') \hat \rho (t) \hat S^{(+)}_i (\omega) - \hat S^{(+)}_i (\omega) \hat S^{(-)}_i (\omega') \hat \rho (t) \mright]     \\
&+ \Gamma_i (\omega) n (\omega, T_i) \mleft[ \hat S^{(-)}_i (\omega') \hat \rho (t) \hat S^{(+)}_i (\omega) - \hat \rho (t)\hat S^{(+)}_i (\omega) \hat S^{(-)}_i (\omega') \mright]   \\
&+ \Gamma_i (\omega) [n (\omega, T_i) + 1] \mleft[ \hat S^{(+)}_i (\omega) \hat \rho (t) \hat S^{(-)}_i (\omega') - \hat S^{(-)}_i (\omega') \hat S^{(+)}_i (\omega) \hat \rho (t) \mright]    \\
&+ \Gamma_i (\omega') [n (\omega', T_i) + 1] \mleft[ \hat S^{(+)}_i (\omega) \hat \rho (t) \hat S^{(-)}_i (\omega') - \hat \rho (t) \hat S^{(-)}_i (\omega') \hat S^{(+)}_i (\omega) \mright]  \\
&+ {\rm \Omega}^{+}_i (T_i) \mleft[ \hat S^{(0)}_i \hat \rho (t) \hat S^{(0)}_i - \hat S^{(0)}_i \hat S^{(0)}_i \hat \rho (t)  \mright]    
 + {\rm \Omega}^{'+}_i (T_i) \mleft[ \hat S^{(0)}_i \hat \rho (t) \hat S^{(0)}_i - \hat \rho (t)\hat S^{(0)}_i (\omega') \hat S^{(0)}_i \mright]   \\
&+ {\rm \Omega}^{-}_i (T_i) \mleft[ \hat S^{(0)}_i \hat \rho (t) \hat S^{(0)}_i - \hat \rho (t)\hat S^{(0)}_i \hat S^{(0)}_i  \mright]     
 + {\rm \Omega}^{'-}_i (T_i) \mleft[ \hat S^{(0)}_i \hat \rho (t) \hat S^{(0)}_i - \hat S^{(0)}_i \hat S^{(0)}_i \hat \rho (t) \mright]
\,\,\,\,\,\,\,\,\,\,
\bigg\} \, ,
\end{split}
\end{equation}
\end{widetext}
where  ${ \Gamma_i(\omega) = 2\pi g_i(\omega)\abs{\alpha_i(\omega)}^2 }$ are the damping rates of the system, ${n(\nu, T_i)}$ thermal populations, and $\Omega^{\pm}_{i}(T)$ are temperature-dependent pure-dephasing rates.
In the expression for ${\cal L}_{\rm gme}\hat\rho$, several terms oscillate at frequencies that can be significantly larger than the characteristic damping rates $\Gamma_I$, because the RWA has not ben applied \cite{settineri2021}.
Notice that in the case of well-separated energy levels, i.e. $|\omega - \omega'| \gg \Gamma_i$, these rapidly oscillating terms average out over time and can be neglected.
Terms oscillating at frequencies $\pm(\omega + \omega')$ arising from products such as $\hat{S}_i^{(\pm)}\hat{S}_i^{(\pm)}$ have already been discarded in \eqref{eq:Lgme1}.  Similarly, terms involving $\hat{S}_i^{(+)}\hat{S}_i^{(0)}$ and $\hat{S}_i^{(-)}\hat{S}_i^{(0)}$, which oscillate 
at $\omega$ and $-\omega'$  respectively, are also neglected due to the same energy scale separation.
The only remaining non-negligible terms are those oscillating at frequencies ${\pm(\omega - \omega')}$ which originate from cross terms like $\hat{S}_i^{(\pm)}\hat{S}_i^{(\mp)}$.
These terms can be comparable to the damping rates and are retained, but modulated by a filter function $F(\omega,\omega')$ which suppresses  off-resonant contributions. The filtered Liouvillian is then expressed as
$${{\cal L}_{\rm gme}^{filt}\hat\rho = {\cal L}_{\rm gme}\hat\rho \cdot F(\omega,\omega')}\,.$$ 
Additionally, there are zero-frequency terms arising from  $\hat{S}_i^{(0)}$ which induce pure dephasing. These dephasing terms are typically negligible at low temperatures. The pure dephasing terms are:
\begin{equation} 
\begin{split}
     &\Omega_i^{'\pm}(T_i) = {\!\!} \int_0^t {\!\!} d\tau\!\!\int_0^{\infty} {\!\!\!\!} d\nu\, g_i(\nu)\abs{\alpha_i(\nu)}^2[n(\nu,T_i)+1]e^{\pm i\nu\tau}  , \\
     &\Omega_i^{\pm}(T_i) = {\!\!} \int_0^t {\!\!} d\tau\!\!\int_0^{\infty} {\!\!\!\!} d\nu\, g_i(\nu)\abs{\alpha_i(\nu)}^2[n(\nu,T_i)]e^{\pm i\nu\tau}\, .
\end{split}
\end{equation}
The system is assumed to interact with an Ohmic bath \cite{Settineri2018dissipation} with damping rates and pure dephasing rates described by the following relations:
\bea \label{eq:app_gammaomega}
    & \Gamma_i (\omega) = 2\pi g_i(\omega)\abs{\alpha_i(\omega)}^2 = \dfrac{\gamma_i}{f_i} \omega\, , \\
    & \Omega_i^{'\pm}(T_i) = \Omega_i^{\pm}(T_i) = \Omega(T_i) = \dfrac{\gamma_i}{4 f_i}T_i \, . \label{eq:app_gammaomega2}
\eea
Here $g_i(\omega)$ denotes the bath density of state, $\alpha(\omega)$ is the coupling strength between the system and the bath, $\omega$ is the bath frequency, and $f_i$ is the bare characteristic frequency of the subsystem under consideration (either the $k$-th individual qubit or the cavity mode).


In this paper, we presented emission spectra considering (i) a collective reservoir for all the emitters, (ii) individual (local) reservoirs for each TLS, or (iii) the coexistence of both types of reservoirs.

For case (i), the dissipation operator is ${\hat s_{\rm coll} = \frac{1}{N}\sum_k \sigma_k^-}$, and the corresponding damping and pure dephasing rates are given by
\begin{equation}
    \Gamma_{\rm{coll}}(\omega) = \frac{\gamma_{\rm{coll}} \, \omega}{f_a}, \quad \Omega^{'\pm}_{\rm{coll}}(T_{\rm{coll}}) = \frac{\gamma_{\rm{coll}}}{4 f_a} T_{\rm{coll}}\, ,
\end{equation}
where $f_a$ is the transition frequency of all the identical TLSs.
For case (ii), the dissipation operators are ${\hat s_{k,\, \rm ind} = \sigma_k^-}$, while the damping and pure dephasing rates are
\begin{equation}
    \Gamma_{k,\,\rm{ind}}(\omega) = \frac{\gamma_{\rm{ind}} \, \omega}{f_k}, \quad \Omega^{'\pm}_{k,\,\rm{ind}}(T_k) = \frac{\gamma_{\rm{ind}}}{4 f_k} T_k\, ,
\end{equation}
where $f_k$ is the transition frequency of the $k$-th TLS.

\newpage
\bibliography{main}

\end{document}